\documentclass[lettersize,journal]{./cls/IEEEtran}
\usepackage{amsmath,amsfonts}
\usepackage{algorithmic}
\usepackage{algorithm}
\usepackage{array}
\usepackage[caption=false,font=normalsize,labelfont=sf,textfont=sf]{subfig}
\usepackage{textcomp}
\usepackage{stfloats}
\usepackage{url}
\usepackage{verbatim}
\usepackage{graphicx}
\usepackage{cite}

\newcommand{\be}{\begin{equation}}
\newcommand{\ee}{\end{equation}}

\def\bal#1\eal{\begin{align}#1\end{align}}
\def\baln#1\ealn{\begin{align*}#1\end{align*}}

\newcommand{\ben}{\begin{equation*}}
\newcommand{\een}{\end{equation*}}

\newcommand{\bbm}{\begin{bmatrix}}
\newcommand{\ebm}{\end{bmatrix}}

\newcommand{\bBm}{\begin{Bmatrix}}
\newcommand{\eBm}{\end{Bmatrix}}

\newcommand{\bvm}{\begin{vmatrix}}
\newcommand{\evm}{\end{vmatrix}}

\newcommand{\bVm}{\begin{Vmatrix}}
\newcommand{\eVm}{\end{Vmatrix}}

\newcommand{\bpm}{\begin{pmatrix}}
\newcommand{\epm}{\end{pmatrix}}

\newcommand{\bnm}{\begin{matrix}}
\newcommand{\enm}{\end{matrix}}

\newcommand{\bi}{\begin{itemize}}
\newcommand{\ei}{\end{itemize}}

\newcommand{\bse}{\begin{subequations}}
\newcommand{\ese}{\end{subequations}}

\newcommand{\bald}{\begin{aligned}}
\newcommand{\eald}{\end{aligned}}

\newcommand{\bxi}{\boldsymbol{\xi}}
\newcommand{\bu}{\boldsymbol{u}}

\newcommand{\bx}{\boldsymbol{x}}

\newcommand{\bv}{\boldsymbol{v}}
\newcommand{\bI}{\boldsymbol{I}}

\newcommand{\bsig}{\boldsymbol{\sigma}}

\newcommand{\R}{\mathbb{R}}

\newcommand{\umax}[1]{{#1}_{max}}

\newtheorem{remark}{Remark}

\newtheorem{prop}{Proposition}
\newtheorem{procedure}{Procedure}

\usepackage{tikz}
\usepackage{physics}
\usepackage{amsmath}
\usepackage{tikz}
\usepackage{mathdots}
\usepackage{yhmath}
\usepackage{cancel}
\usepackage{color}
\usepackage{siunitx}
\usepackage{array}
\usepackage{multirow}
\usepackage{amssymb}
\usepackage{gensymb}
\usepackage{tabularx}
\usepackage{extarrows}
\usepackage{booktabs}
\usetikzlibrary{fadings}
\usetikzlibrary{patterns}
\usetikzlibrary{shadows.blur}
\usetikzlibrary{shapes}

\usepackage{pgfplots}
\pgfplotsset{width=10cm,compat=1.9}

\usepackage{cases} 
\usepackage[hidelinks]{hyperref} 
\hypersetup{
    pdftitle={Saturated Control for a quadcopter HTD\_FB\_IP}
    }

\newcommand{\satlambda}{\text{sat}_\lambda (-\gamma B^\top P\bxi)}
\newcommand{\brho}{\mathcal{B}(\rho)}
\newcommand{\reff}[1]{#1^{\mathrm{ref}}}
\newcommand{\BPep}{\mathcal{B}_P(\varepsilon)}
\newcommand{\QEDA}{\hfill\ensuremath{\square}}

\newcommand{\videourl}{https://youtu.be/cgXNKBvSBRM}

\begin{document}

\title{
A flatness-based saturated controller design \\for a quadcopter with experimental validation
}

\author{Huu-Thinh Do$^*$, Franco Blanchini$^{**}$, Ionela Prodan$^*$
\thanks{$^{*}${Univ. Grenoble Alpes, Grenoble INP$^\dagger$, LCIS, 26000 Valence, France}.
Email: \textsf{\{huu-thinh.do,ionela.prodan\}@lcis.grenoble-inp.fr}
\newline
$^\dagger$Institute of Engineering and Management Univ. Grenoble Alpes. 
}
\thanks{$^{**}$Dipartimento di Matematica e Informatica, Universit\`a di Udine, 33100 Udine, Italy. Email: \textsf{franco.blanchini@uniud.it}}
}

\maketitle
\begin{abstract}
Using the properties of differential flatness, a controllable system, such as a quadcoper model, may be transformed into a linear equivalent system via a coordinate change and an input mapping. This is a straightforward advantage for the quadcopter’s controller design and its real-time implementation. However, one significant hindrance is
that, while the dynamics become linear in the new coordinates (the flat output space), the input constraints become convoluted.  This paper addresses an explicit pre-stabilization based control scheme which handles the input constraints for the quadcopter in the flat output space with a saturation component. The system's stability is shown to hold by Lyapunov-stability arguments. Moreover, the practical viability of the proposed method is validated both in simulation and experiments over a nano-drone platform. Hence, the flatness-based saturated controller not only ensures stability and constraints satisfaction, but also requires very low computational effort, allowing for embedded implementations.
\end{abstract}

\begin{IEEEkeywords} Quadcopter, differential flatness, constraint-handling controller, pre-stabilization control design, ellipsoidal invariant sets, saturated control.
\end{IEEEkeywords}

\section{Introduction}
Of all the unmanned aerial vehicles, multicopters have drawn remarkable attention in both research and practical use thanks to their ability to accomplish vertical take-off and landing as well as static hovering, which are applied in various logistic, surveillance or law-enforcing applications\cite{huang2017structure,liu2018novel,zheng2021multilayer}.
However, although being investigated for decades\cite{hua2009control,santoso2019hybrid,nguyen2020flat}, 
their guidance, navigation and control
remain problematic due to the model's nonlinearity and physical constraints.

Typically, to deal with the nonlinearity, the dynamics are linearly approximated via the Taylor expansion \textit{near} its equilibrium points and then governed with classic or modern control approaches (e.g. fuzzy logic, linear quadratic regulator)\cite{santoso2019hybrid,garcia2017modeling,Santoso8370894}. 
Although the such control synthesis is proven to be generic and effective, one possible downside is its sensitivity to uncertainty, which is caused by the approximation-based inexact dynamics.
Moreover, theoretically, the closed-loop stability proof is often given only for the approximated system, in lieu of the original nonlinear one \cite{santoso2019hybrid,garcia2017modeling}, hence neglecting the entire left-over remainder of the approximation. 

On the contrary, as an intrinsic property of the quadcopter, the system's dynamics is known to be \textit{differentially flat} \cite{levine2009analysis,formentin2011flatness,nguyen2020flat}. 
Namely, all the system's states and inputs can be algebraically expressed in terms of a special output, called \textit{flat output}, and a finite number of its derivatives.
Consequently, thanks to this input-output relationship, the system's  motion can be exactly described by finite chains of trivial integrators after a diffeomorphism (bijective differentiable coordinate change) and a dynamic feedback linearization law.
In other words, the system can be equivalently represented by a linear controllable system in new coordinates, called the \textit{flat output space}. Hence, the control problem is now solved by closing the loop for such a linear system, converting back the control input to the original description and applying it to the real system. This approximation-free approach undeniably simplifies the system description, yet, its corresponding constraints are convoluted (in general, becoming nonlinear).
To overcome this hindrance, several approaches have been employed, including stabilizing the linear model by conservatively sketching the new constraints by its box-type subsets or governing the vehicle via dynamically constrained feedforward reference \cite{lu2017constrained,mueller2013model,sun2022comparative}. 
However, the solutions given by 
these aforementioned proposals or approaches are either computationally burdensome or conservatively addressed. 

Therefore, in this work, we present an in-depth investigation of the quadcopter representation in its flat output space together with the distorted (by mapping in the flat output space) constraint description. Then, based on the quadcopter's characterization in the flat output space, we proposed a novel flatness-based saturated control design (FBSC) guaranteeing both stability and constraint satisfaction.
Briefly, the salient contributions of our work are summarized as follows. We:
\begin{itemize}
    \item investigate the nonlinear representation of the  quadcopter's input constraints in the flat output space, which to the best of our knowledge, has not been thoroughly studied in the literature (conservative approximations were usually employed \cite{mueller2013model,nguyen2020flat,nguyen2020stability,liu2018novel});
    \item design an optimization-based saturated controller which ensures both stability and constraint satisfaction, based on the linearized dynamics in the new coordinate and its associated nonlinear constraints;
    \item propose a computational low-cost explicit procedure to implement the proposed algorithm;
    \item validate the benefits of the proposed controller through simulation and experimental tests by using the Crazyflie 2.1 nano-drone platform \cite{giernacki2017crazyflie}. The video for the experiment is available at \url{\videourl}.
\end{itemize}
\IEEEpubidadjcol 

The remainder of the paper is structured as follows.
Section \ref{sec:Preliminaries} presents the quadcopter model, its flat characterization
 and operational constraints. 
 Furthermore, invariance and Lyapunov-based tools are briefly introduced, which will prove instrumental for the proposed controller stability analysis.
Section \ref{sec:Explicit} presents the saturated control design together with theoretical proofs for constraint satisfaction and stability. Section \ref{sec:experiments} shows the advantages of the proposed controller through simulations and experiments under different trajectory scenarios. Comparisons and discussions highlight the strengths of our novel flatness-based saturated controller design. Finally, Section \ref{sec:conclude} summarizes and provides insights on future work.

\emph{Notations:}
Let us denote upper-case letters as matrices with suitable dimension. Next, $\bI_n$ and $\boldsymbol 0_n$ denote an $n\times n$ identity and zero square matrix, respectively. Similarly, $ \boldsymbol 0_{a\times b}$, represents a matrix of dimension $a\times b$ all of whose components are zero. $\text{diag}(\cdot)$ generates a diagonal matrix created by the employed components. Next, vectors are represented by bold letters (e.g, $\bv,\bx$). $\|\bv\|_Q$ and $\|\bv\|_2$ denote the weighted norm $\sqrt{\bv^\top Q \bv}$ and the Euclidean norm $\sqrt{\bv^\top  \bv}$, respectively. Then, $\mathcal{B}(r),\mathcal{B}_M(\varepsilon)\subset\R^n$ defines the balls characterized by $\mathcal{B}(r)=\{\bx\in\R^n:\|\bx\|_2^2 \leq r\}$ and $\mathcal{B}_M(\varepsilon)=\{\bx\in\R^n:\|\bx\|_M^2 \leq \varepsilon\}$, respectively.
Next, with $\bx=[x_1,...,x_n]^\top\in\R^n$, the expression $\bx\leq 0$ implies the set of componentwise inequalities $x_i\leq 0,i\in\{1,...,n\}$. $M\succ0$ and $M\succeq 0$ imply that $M$ is positive definite and semi-definite, respectively. Similarly, $M\prec 0$ and $M\preceq 0$ imply $-M\succ 0$ and $-M\succeq 0$, respectively. For a set $\mathcal{X}$, $\partial \mathcal{X} $ denotes its boundary. Finally, the superscript ``$\mathrm{ref}$" denotes the desired reference signal for the system to track (e.g., $\reff\bu$).

\section{Model representation and prerequisites}
\label{sec:Preliminaries}
\begin{figure}[htbp]
    \centering
    \resizebox{0.48\textwidth}{!}{\input{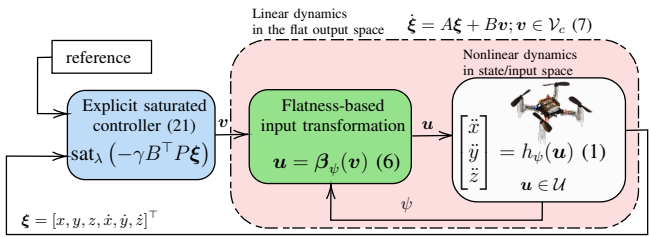}}
    \caption{Flatness-based control scheme for the quadcopter.}
    \label{fig:scheme_drone}
\end{figure}

Let us first provide an overview in Fig. \ref{fig:scheme_drone} of the general flatness-based control scheme we adopt in this paper. 
One of the contributions resides in the fact that the
control synthesis (the blue block in Fig. \ref{fig:scheme_drone}) is done in the flat output space (the red block) where the dynamics is linear.
The
reference for the linearized dynamics together with the feedback signals are given to a novel saturated (the blue block) to compute the stabilizing control action, $\bv$. Then, such input, $\bv$, in the flat output space, is transformed back to the real input, $\bu$, via a flatness-based transformation (the green block). This input $\bu$ is, in turn, applied to the quadcopter  (the white block in Fig. \ref{fig:scheme_drone}), closing the loop for the entire scheme.

In this section, 
we refer to the white and green blocks of Fig. \ref{fig:scheme_drone} (with a particular attention to the constraints' representation in the flat output space), while in Section \ref{sec:Explicit} we will handle the blue block. Hence, in the following, we
first introduce the quadcopter model
characterization 
in the flat output space. 
Then, we introduce some invariance and Lyapunov-based tools which will be employed.

\subsection{Quadcopter's model and its characterization in the flat output space}
The translational dynamics of a quadcopter is described as:
 \be 
 \begin{aligned}
	\bbm\ddot x \\ \ddot y \\ \ddot z \ebm&=\bbm 
		T(\cos \phi \sin \theta \cos \psi+\sin \phi \sin \psi) \\
		T(\cos \phi \sin \theta \sin \psi-\sin \phi \cos \psi) \\
		T\cos \phi \cos \theta -g
	\ebm
	\triangleq h_\psi(\bu),
\end{aligned}
	\label{eq:drone_dyna}
 \ee 
where $x,y,z$ represent the positions of the quadcopter in the global frame's three axis. $\psi$ is the yaw angle, which is considered as a known parameter for this system, $g$ is the gravitational acceleration while $\bu\in \R^3$ collects the three inputs of the system $\bu=[T,\;\phi,\; \theta]^\top $, including, respectively, the thrust, roll and pitch angles. The input $\bu$ will be
is actuated by the low level controller that
imposes proper speed values to the four propellers.
Finally, let us denote $\mathcal{U}\subset \R^3$ the constraint set for the input $\bu\in \R^3$, which is described as:
\be 
\mathcal{U}=\left\{ \bu:0\leq T\leq T_{max}, |\phi|\leq \phi_{max},|\theta|\leq \theta_{max}\right\}
\label{eq:orginal_input_constr}
\ee
where $\umax T>0,  (\umax \phi ,\umax \theta)\in (0;\pi/2)^2$ are constant bounds of the inputs.

The model \eqref{eq:drone_dyna} for thrust-propelled system is known to be differentially flat \cite{nguyen2017reliable,sun2022comparative}. Namely, all state and input variables can be differentially parameterized by a fictitious output, called the \textit{flat output}, and a finite number of its derivatives. Indeed, by choosing the flat output $\bsig$ as:
\be 
\bsig=[\sigma_1,\sigma_2,\sigma_3]^\top\triangleq[x,y,z]^\top\in\R^3,
\label{eq:flat_output_def}
\ee
we achieve the following \textit{flat representation}:
\begin{subequations}
	\begin{align}
x&=\sigma_1, y=\sigma_2, z=\sigma_3,\label{eq:flat_rep_quada}  \\
T&=\sqrt{\ddot\sigma_1^2+\ddot\sigma_2^2+(\ddot\sigma_3+g)^2},\label{eq:flat_rep_quadb} \\
\phi&=\arcsin{\left({(\ddot\sigma_1\sin{\psi}-\ddot\sigma_2\cos{\psi})}/{T}\right)},\label{eq:flat_rep_quadc} \\
\theta&=\arctan{\left({(\ddot\sigma_1\cos{\psi}+\ddot\sigma_2\sin{\psi})}/{(\ddot\sigma_3+g)}\right)}.\label{eq:flat_rep_quadd}
	\end{align}
	\label{eq:flat_rep_quad}
\end{subequations}
Then by exploiting \eqref{eq:flat_rep_quadb}-\eqref{eq:flat_rep_quadd},
the linearizing feedback can be deduced as:
\begin{subequations}
	\begin{align}
T&=\sqrt{v_1^2+v_2^2+(v_3+g)^2},\label{eq:linearization_a} \\
\phi&=\arcsin{\left({(v_1\sin{\psi}-v_2\cos{\psi})}/{T}\right)},\label{eq:linearization_b} \\
\theta&=\arctan{\left({(v_1\cos{\psi}+v_2\sin{\psi})}/{(v_3+g)}\right)},\label{eq:linearization_c}
	\end{align}
\label{eq:linearization}
\end{subequations}
or in a more compact form:
\be 
\bu =\boldsymbol\beta_\psi(\bv),
\label{eq:linearization_compact}
\ee 
where $\bv=[v_1,v_2,v_3]^\top\in\R^3$ is the new input of the transformed system.
Then, under the condition $v_3\geq-g$ and the mapping \eqref{eq:linearization_compact}, the system \eqref{eq:drone_dyna} becomes:
\be 
    \dot\bxi =A\bxi + B\bv ,
\label{eq:linearized_drone}
\ee 
with $\bxi=[\bsig^\top,\dot\bsig^\top]^\top=[x,y,z,\dot x,\dot y, \dot z]^\top\in\R^6$ and the matrices $A=\bbm \boldsymbol{0}_{3}& \bI_{3} \\
\boldsymbol{0}_{3}&\boldsymbol{0}_{3}\ebm , B=\bbm \boldsymbol{0}_{3}\\ \bI_{3}\ebm $.


With the dynamics \eqref{eq:linearized_drone} and the feedback law \eqref{eq:linearization}, in the next subsection, we analyze the constraints on $\bv$ to provide a background for the controller design developed later.

\vspace{-0.5cm}
\subsection{Input constraints description in the flat output space}
We denote the constraint set for the input $\bv$ in \eqref{eq:linearized_drone} as:
\be 
\mathcal{V}=\{\bv\in\R^3:\boldsymbol\beta_\psi(\bv)\in\mathcal{U} \text{ as in \eqref{eq:orginal_input_constr}}\}.
\label{eq:nonconvex_set_flat}
\ee 
Next, we show by a counterexample that $\mathcal{V}$ as in \eqref{eq:nonconvex_set_flat} is a non-convex set. More specifically, $\mathcal{V}$ contains the two points $\bv_\pm =h_\psi([\umax T, \pm \umax \phi,\pm \umax \theta]^\top)$ but it does not contain their midpoint $0.5(\bv_-+\bv_+)$. Thus, this hinders us from embedding the control \eqref{eq:linearized_drone} into any convex optimization based framework (e.g, model predictive control). Second, the description of $\mathcal{V}$ requires a constant update of the yaw angle $\psi$, which is usually time-varying in practical applications. This requirement 
is very demanding in terms of computation.
For those reasons, $\mathcal{V}$ is rarely exploited in the literature. Indeed, as an alternative, a $\psi$-free and convex subset of $\mathcal{V}$, denoted as $\mathcal{V}_c$, is frequently employed, (as in \cite{nguyen2018effective,nguyen2020flat}):
\begin{align}
    \mathcal{V}_c=\Big\{&\bv\in \R^3:  \bbm {v_1^2+v_2^2+(v_3+g)^2} -T_{max}^2\\
    v_1^2+v_2^2- (v_3^2+g)^2 \tan^2\epsilon_{max} 
    \ebm    \leq 0, \nonumber\\ 
    &\epsilon_{max}\triangleq\min(\theta_{max},\phi_{max}) \text{ and } v_3\geq-g\Big\}.
    \label{eq:constraint_convex}
\end{align}
\vspace{-0.925cm}
\begin{figure}[htbp]
    \centering
    \includegraphics[scale=0.6]{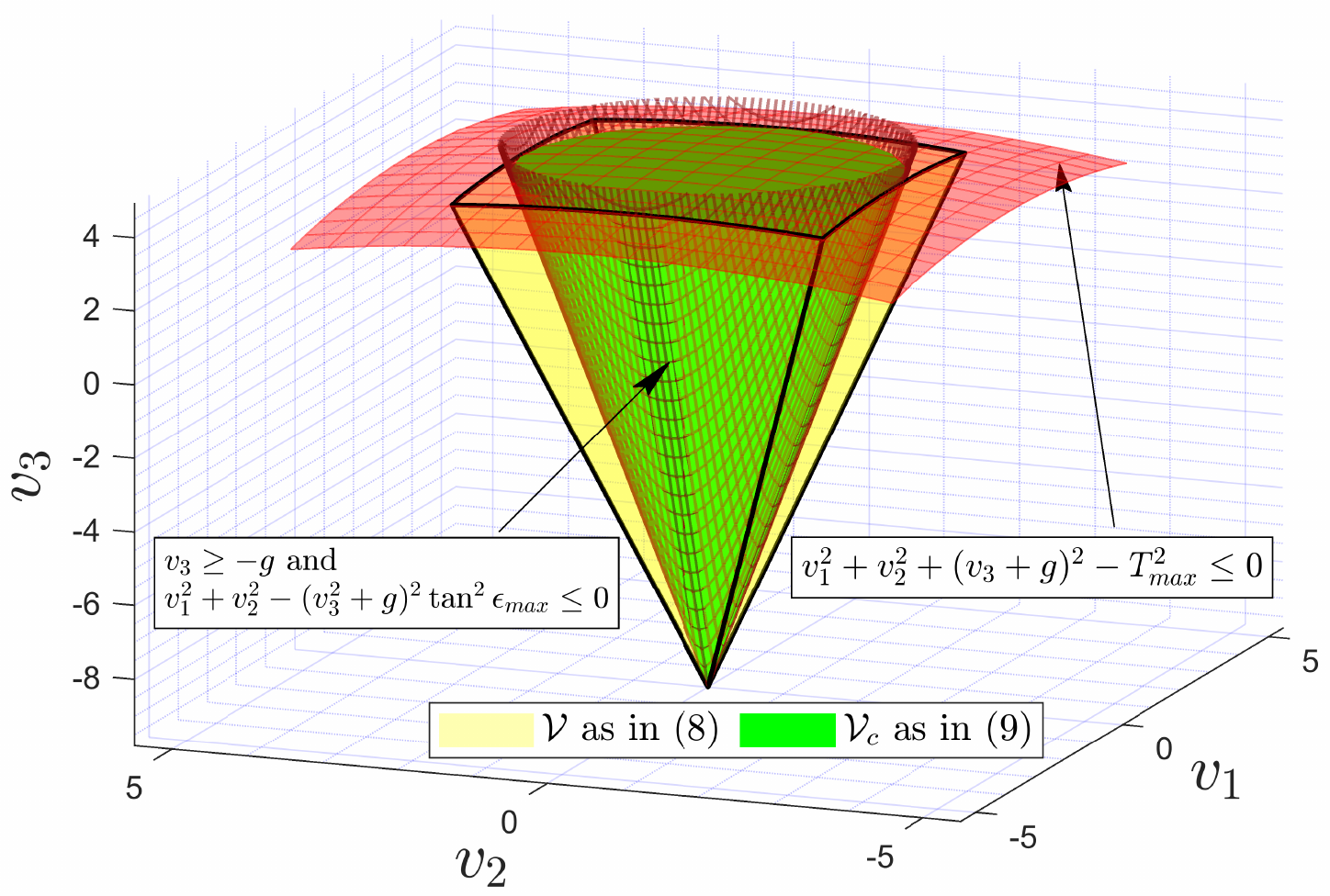}
    \caption{Quadcopter's input constraints in the flat output space with $\psi=0\;(rad)$, $g=\umax T/1.45= 9.81\; (m/s^2)$, $ \umax \phi=\umax\theta=\pi/18 \;(rad)$).}
    \label{fig:Twosets_two_spaces}
\end{figure}

In detail, the containment of $\mathcal{V}_c$ in \eqref{eq:constraint_convex}, inside $\mathcal{V}$ as in \eqref{eq:nonconvex_set_flat} can be shown by considering the following inequalities  (see Fig. \ref{fig:Twosets_two_spaces}). Firstly, for the constrained angles, by applying Cauchy-Schwarz inequality in \eqref{eq:linearization_b} and \eqref{eq:linearization_c}, we always have\cite{nguyen2018effective}:
\be
|\sin\phi|\leq \sin\epsilon(\bv)\text{ and }|\tan\theta|\leq \tan\epsilon(\bv),
\ee 
where $\epsilon(\bv)\triangleq \arctan(\sqrt{{(v_1^2+v_2^2)}/{(v_3+g)^2}})$ interprets as a $\bv$-dependent angle upper bounding the roll $\phi$ and the pitch angle $\theta$. Then, with the assumption $|\theta|,|\phi|\leq \pi/2$ as in \eqref{eq:orginal_input_constr} and by imposing $\epsilon(\bv)\leq\epsilon_{max}\triangleq\min(\theta_{max},\phi_{max})$, we can ensure $[|\theta|-\theta_{max},|\phi|-\phi_{max}]^\top\leq 0$ thanks to the local monotonicity of $\tan(\cdot)$ and $\sin(\cdot)$ functions. 
Secondly, for the constrained thrust $T$, it is straightforward that if $v_1^2+v_2^2+(v_3+g)^2 -T_{max}^2\leq 0$ holds, we have $0\leq T\leq \umax T$. Therefore, $\mathcal{V}_c$ is a subset of $\mathcal{V}$. Moreover, the convexity of $\mathcal{V}_c$ can be proven by showing that $\mathcal{V}_c$ is the intersection of two convex sets: a ball described by $v_1^2+v_2^2+(v_3+g)^2 -T_{max}^2\leq 0$ and a convex cone formed by two inequalities: $v_3\geq -g$ and $v_1^2+v_2^2- (v_3^2+g)^2 \tan^2\epsilon_{max}$. An illustration of $\mathcal{V}$ and $\mathcal{V}_c$ is given in Fig. \ref{fig:Twosets_two_spaces}
.

Our problem has now been reduced
to the control of the quadcopter with 
 a linear dynamics \eqref{eq:linearized_drone} subject to a non-linear (but convex) constraint, $\mathcal{V}_c$ as in \eqref{eq:constraint_convex}
 in the flat output space 
 (the blue block in Fig. \ref{fig:scheme_drone}), which must ensure the constraint satisfaction and stability. Then,
 the real input $\bu$ is computed via the mapping $\bu=\boldsymbol\beta_\psi(\bv)$ given in \eqref{eq:linearization_compact} (the green block in Fig. \ref{fig:scheme_drone}). That input $\bu$ then will be implemented on the original system \eqref{eq:drone_dyna} assuming that the state is measured.


\subsection{Invariance and Lyapunov-based tools}
In this subsection, we recall some theoretical notions that will be adopted in section III.

We will use ellipsoidal sets to define a domain of attraction for our closed-loop system, so we remind the condition of invariance for an ellipsoidal set.

\begin{prop}
\label{prop:Nagumo}
\textit{(Nagumo's invariance principle \cite{blanchini2008set}) }
Consider a dynamical system described as:
\be
\dot\bxi=f(\bxi,\bv),
\label{eq:general_dyna_sys}
\ee 
with $\bxi\in\R^n,\bv\in\R^m$ representing the state and the input vector, respectively. Then, the ellipsoid set $\mathcal{E}=\{\bxi\in\R^n:\bxi^\top P\bxi \leq \varepsilon\}$ with $P\succ 0, \varepsilon>0$ is an invariant set associated with the dynamics \eqref{eq:general_dyna_sys} if and only if:
\be 
\bxi_b^\top P f(\bxi_b,\bv)\leq 0,\, \forall \bxi_b\in\partial\mathcal{E}.
\label{eq:nagumo_general}
\ee 
where $\bxi_b\in\partial\mathcal{E}$ denotes the boundary points of the set $\mathcal{E}$.
\QEDA
\end{prop}

Let us recall the gradient-based control, one of the most classical controllers to stabilize a linear system based on a quadratic Lyapunov function.

\begin{prop} (\textit{Gradient-based control with a quadratic control Lyapunov function\cite{blanchini2008set}.})
\label{prop:GB_lya}
Consider the following linear time-invariant (LTI) system
\be 
\dot \bxi = A\bxi +B\bv
\label{eq:general_LTI}
\ee 
with $\bxi\in\R^n,\bv\in\R^m$. Then, for any $\alpha>0$ and $\gamma\geq 1$, by imposing a gradient-based control:
\be
\bv = -\gamma B^\top P\bxi,
\label{eq:general_GB}
\ee 
satisfying the condition of $P\succ 0$ and:
\be 
(A-BB^\top P)^\top P + P(A-BB^\top P) \preceq -\alpha P,
\label{eq:stabilizing_condition}
\ee 
then the quadratic positive definite function: 
\be V=\bxi^\top P \bxi, 
\label{eq:general_lya_func}
\ee 
is a control Lyapunov function for the dynamics \eqref{eq:general_LTI} associated with the control \eqref{eq:general_GB}. As a result, this control setup ensures the exponential convergence of the system to the equilibrium point $\bxi=\boldsymbol 0_{n\times 1}, \bv=\boldsymbol 0_{m\times 1}$. 
\QEDA
\end{prop}

\begin{IEEEproof}
Justification for Proposition \ref{prop:GB_lya} can be shown by computing the time derivative of $V$ in \eqref{eq:general_lya_func}. It is trivial to achieve, $\forall \gamma\geq 1$:
\be 
\begin{aligned}
    \Dot{V}&=2\bxi^\top P(A\bxi+B\bv)\leq-\alpha V - 2(\gamma-1)\|B^\top P\bxi\|_2^2,\\
    &\leq -\alpha V,
\end{aligned}
\label{eq:lyapunov_gen_nosat}
\ee 
Hence $\forall\gamma \geq 1$, the dynamics \eqref{eq:general_LTI} is stabilized by the feedback \eqref{eq:general_GB} and the system converges to $\bxi = \boldsymbol 0_{n\times 1},\bv=\boldsymbol 0_{m\times 1}$.
\end{IEEEproof}

We recall the $\mathcal{S}$-procedure transforming quadratic inequalities to linear matrix inequalities (LMI).
\begin{prop}
\label{prop:Sproc}
($\mathcal{S}$-\textit{procedure}\cite{boyd1994linear}) With $\bxi\in\R^n$, consider the two following scalar-valued quadratic functions:
\be 
\mathcal{F}_i(\bxi)=\bxi^\top Q_i \bxi +2\boldsymbol\mu_i^\top\bxi+w_i,\,i\in\{1,2\}
.
\ee 
Then, with $Q_i=Q_i^\top$, the two following statements are equivalent:
\begin{enumerate}
    \item $\mathcal{F}_1(\bxi) \geq 0 \Rightarrow\mathcal{F}_2(\bxi)\geq 0$.
    \item $\exists\tau\geq0\in\R$ such that:
    \be 
    \bbm 
    Q_1 & \boldsymbol\mu_1 \\ \boldsymbol\mu_1^\top & w_1
    \ebm-
    \tau\bbm 
    Q_2 & \boldsymbol\mu_2 \\ \boldsymbol\mu_2^\top & w_2
    \ebm \succeq 0.
    \ee 
\end{enumerate}
\QEDA
\end{prop}

Considering the above tools, in the subsequent section, we introduce a specialized saturated controller for the constraint set $\mathcal{V}_c$ together with the stability and constraint satisfaction guarantee.

\section{Explicit saturated control design}
\label{sec:Explicit}
In this section, we propose a novel optimization-based saturated controller 
that provides noteworthy
advantages for real-time embedded implementation.
\subsection{Controller characterization}
To proceed with a Lyapunov-based design, we recall the control problem defined in the flat output space, where the goal is to stabilize the system \eqref{eq:linearized_drone} under constraints \eqref{eq:constraint_convex}.
\vspace{-0.01cm}

The following proposition describes the proposed controller with stability guarantees.
\begin{prop}
Consider a ball
 \be 
 \mathcal{B}(\rho)=\{\bv\in\R^3:\|\bv\|_2^2\leq \rho\} \text{ s.t } \mathcal{B}(\rho)\subseteq \mathcal{V}_c \text{ as in \eqref{eq:constraint_convex}},
 \label{eq:Ball_in_Vc}
 \ee 
some scalars $\alpha>0, \gamma\geq 1$, a matrix $P\succ 0$ satisfying the condition \eqref{eq:stabilizing_condition}, and the saturated controller defined as:
\be 
\bv = \text{sat}_\lambda (-\gamma B^\top P\bxi),
\label{eq:control_sat}
\ee 
with $\text{sat}_\lambda (\bv)$ given by:
\begin{subequations}
    \begin{align}
         \text{sat}_\lambda (\bv)& =
         \begin{cases}
             \bv \text{ if } \bv\in \mathcal{V}_c,&\\
 \lambda^*(\bv)\bv \text{ if } \bv\notin \mathcal{V}_c
         \end{cases}
\label{eq:def_sat}\\
\text{with } \lambda^*(\bv)&=\underset{{
\lambda \bv \in \mathcal{V}_c}}{\arg \max }\;\lambda .
\label{eq:def_sat_opt}
    \end{align}
    \label{eq:def_sat_full}
\end{subequations}
The following results hold for the system \eqref{eq:linearized_drone} under the controller \eqref{eq:control_sat}:
\begin{enumerate}
    \item The control action $\bv$ as in \eqref{eq:control_sat} is contained in $\mathcal{V}_c$, i.e.,
\be \bv=\satlambda\in\mathcal{V}_c\,\forall \bxi \in\R^6, \gamma\geq 1.
\label{eq:constraint_satisfaction}
\ee 
\item The following ellipsoidal set is rendered invariant:
\be 
\mathcal{B}_P(\varepsilon)=\{\bxi\in\R^6:\|\bxi\|_P^2\leq \varepsilon\},
\label{eq:ellipsoid_PI}
\ee 
where the largest $\varepsilon>0$ is found by solving:
\begin{subequations}
\label{eq:find_ellipsoid}
\begin{align}
&\varepsilon^*=\underset{\varepsilon}{\arg \max }\;(\varepsilon ) 
\label{eq:find_ellipsoid_a}\\
\text{s.t }
\bbm
P  & \boldsymbol 0_{6 \times 1} \\ \boldsymbol 0_{1 \times 6} & -\varepsilon
\ebm 
-&\tau
\bbm
P  B  B^\top P  & \boldsymbol 0_{6 \times 1} \\ \boldsymbol  0_{1 \times 6} & -\rho 
\ebm \succeq 0
, \tau \geq 0.\label{eq:find_ellipsoid_b}
\end{align}
\end{subequations}
Moreover, the system \eqref{eq:linearized_drone} 
achieves exponential stability 
with domain of attraction
$\mathcal{B}_P(\varepsilon)$ as in \eqref{eq:ellipsoid_PI}.
\end{enumerate}

\QEDA
\end{prop}
\begin{IEEEproof} 
Constraint satisfaction follows immediately from conditions \eqref{eq:def_sat}-\eqref{eq:def_sat_opt}.

To prove stability, we show that the quadratic function associated with $P$ is decreasing
(its derivative is negative)  
inside  $\mathcal{B}_P(\varepsilon)$. Then, the set  $\mathcal{B}_P(\varepsilon)$ is indeed positively invariant. Hence, overall, once the system is inside the set, its stability and constraints are respected.

Firstly, let us proceed to prove the stability of the system inside $\mathcal{B}_P(\varepsilon)$ by explaining the choice of $\varepsilon$ as in \eqref{eq:find_ellipsoid}. More specifically, by solving \eqref{eq:find_ellipsoid_b}, it can be shown that the containment of the state $\bxi$ inside the ellipsoid $\mathcal{B}_P(\varepsilon)$ implies that $B^\top P \bxi \in \mathcal{B}(\rho)$ as in \eqref{eq:Ball_in_Vc}, i,e:
\be 
\bxi \in \mathcal{B}_P(\varepsilon) \Rightarrow B^\top P \bxi \in \mathcal{B}(\rho).
\label{eq:ellipsoid_condition_constrained_v}
\ee 
Indeed, rewriting \eqref{eq:ellipsoid_condition_constrained_v} explicitly, we have:
\be 
\|\bxi\|_P^2 \leq \varepsilon \Rightarrow \|B^\top P \bxi\|_2^2 \leq \rho
\label{eq:quadratic_ineq}
\ee 
Then by applying the $\mathcal{S}$-procedure, as in Proposition \ref{prop:Sproc}, the condition is  \eqref{eq:quadratic_ineq} equivalently rewritten as:
\be 
\exists\tau\geq 0 \text{ s.t:}
\bbm
P  & \boldsymbol 0_{6 \times 1} \\ \boldsymbol 0_{1 \times 6} & -\varepsilon
\ebm 
-\tau
\bbm
P  B  B^\top P  & \boldsymbol 0_{6 \times 1} \\ \boldsymbol  0_{1 \times 6} & -\rho 
\ebm \succeq 0.
\label{eq:S_procedure_drone}
\ee 
Therefore, by solving \eqref{eq:find_ellipsoid} (which satisfies \eqref{eq:S_procedure_drone}), we can ensure the conditions \eqref{eq:ellipsoid_condition_constrained_v}-\eqref{eq:quadratic_ineq}. 
Consequently, together with the definition \eqref{eq:def_sat} and \eqref{eq:def_sat_opt}, we can state:
\be
\|\lambda^*(\bv)\bv \|_2^2\geq \rho\geq  \|B^\top P \bxi\|_2^2\;\, \forall \bxi \in \mathcal{B}_P(\varepsilon), \bv\in\R^3. 
\label{eq:condition_lambdastart}
\ee

This property, 
hereinafter, serves as an ingredient to show the non-positivity of the Lyapunov function's derivative inside the set $\mathcal{B}_P(\varepsilon)$ as in \eqref{eq:ellipsoid_PI}. 

More specifically, 
considering the fact that $P$ satisfies the condition \eqref{eq:stabilizing_condition},
we can analyze the time-derivative of
the Lyapunov function $V(\bxi)=\bxi^\top P\bxi $ associated with the controller \eqref{eq:control_sat} as:
\be 
\dot V(\bxi) = 2\bxi^\top P (A\bxi+B\satlambda).
\label{eq:lyapunov_sat}
\ee 
Then, according to the definition of the saturation function in \eqref{eq:def_sat_full}, let us inspect the two cases determining the closed-loop behavior
as follows.

\underline{\emph{Unsaturated input:}} In this case, inside $\mathcal{B}_P(\varepsilon)$, we have:
\be 
\bv =\satlambda = -\gamma B^\top P\bxi.
\ee 
Thus, the stability is evident, since the equation \eqref{eq:lyapunov_sat} now becomes identical as \eqref{eq:lyapunov_gen_nosat}, or:
\be 
\dot{V}(\bxi)\leq- \alpha V(\bxi)\leq- \alpha\varepsilon,\;\forall \bxi\in \mathcal{B}_P(\varepsilon).
\label{eq:speed1}
\ee 

\underline{\emph{Saturated input:}} In this case, the controller \eqref{eq:control_sat} yields:
\be 
\bv = \satlambda = -\lambda^*(-\gamma B^\top P\bxi)\gamma B^\top P\bxi.
\label{eq:unsat_control}
\ee 
Then, taking into account \eqref{eq:stabilizing_condition}, equation \eqref{eq:lyapunov_sat} becomes:
\be 
\begin{aligned}
    &\dot V(\bxi) = 2\bxi^\top PA\bxi +2\bxi^\top PB\satlambda,\\
    &\leq \bxi^\top(-\alpha P+2PB B^\top P)\bxi+2\bxi^\top PB\satlambda, \\ 
&\leq -\alpha \|\bxi\|_P^2-2\|B^\top P\bxi\|_2^2(\gamma\lambda^*(-\gamma B^\top P\bxi)-1).
\end{aligned}
\label{eq:lyapunov_saturated}
\ee 
Meanwhile, by replacing $\bv = -\gamma B^\top P\bxi$ into the condition \eqref{eq:condition_lambdastart}, we achieve:
\be
\gamma\lambda^*( -\gamma B^\top P\bxi)\geq 1\;\; \forall\bxi \in \mathcal{B}_P(\varepsilon).
\label{eq:stable_feature}
\ee
As a consequence, \eqref{eq:lyapunov_saturated} yields:
\be 
\dot V(\bxi)\leq -\alpha \|\bxi\|_P^2\leq-\alpha \varepsilon\;\;\forall \bxi\in \mathcal{B}_P(\varepsilon).
\label{eq:speed2}
\ee 

Secondly, the Nagumo's invariance condition for the set $\mathcal{B}_P(\varepsilon)$ can be evidently  proven since, particularly in this case, the condition \eqref{eq:nagumo_general} yields:
\be 
\bxi_b^\top P (A\bxi_b+B \text{sat}_\lambda (-\gamma B^\top P\bxi_b))=\dot V(\bxi_b)/2\leq 0,
\label{eq:nagumo_linear}
\ee 
which holds $\forall\,\bxi_b \in\partial \mathcal{B}_P(\varepsilon)$ because $\dot V(\bxi)<0 \forall \bxi \in \mathcal{B}_P(\varepsilon)$ as proven previously for the stability.
Furthermore, condition \eqref{eq:stable_feature} along the invariance property ensures asymptotic stability with domain of attraction
$\mathcal{B}_P(\varepsilon)$.
\end{IEEEproof}

Following the above demonstration, we provide a brief procedure for the  controller \eqref{eq:control_sat} synthesis.
\vspace{0.075cm}

\fbox{\begin{minipage}{0.445\textwidth}
\begin{procedure}
\label{proc:synthesis_control}
\begin{enumerate}
    \item Choose $\rho>0$, such that $\mathcal{B}(\rho)\subset\mathcal{V}_c$ as in \eqref{eq:Ball_in_Vc};
    \item Choose $\alpha>0$ and solve \eqref{eq:stabilizing_condition} for a stabilizing symmetric matrix $P$;
    \item With any choice $\gamma\geq 1$, the ellipsoid $\mathcal{B}(\varepsilon)$ as in \eqref{eq:ellipsoid_PI} is guaranteed to be invariant.
\end{enumerate}
\end{procedure}
\end{minipage}}
\vspace{0.05cm}

\begin{remark}
It is apparent that no eigenvalue of matrix $A$ in \eqref{eq:linearized_drone} has positive real part, and the pair $(A,B)$ is stabilizable. Hence, as stated in \cite{sontag1990nonlinear,saberi2012control,hu2001control}, for any fixed value of $\rho>0$, there always exists a control law which makes the equilibrium points $(\bxi=\boldsymbol 0_{6\times 1},\bv=\boldsymbol 0_{6\times 1})$ \textit{globally asymptotically stable}. Thus, with a proper choice of $\alpha$, one can enlarge the invariant set $\mathcal{B}_P(\varepsilon)$ defined in \eqref{eq:ellipsoid_PI} arbitrarily just with the nominal controller satisfying:
\be 
\bv=B^\top P\bxi \in \mathcal{B}(\rho) \forall \bxi\in \mathcal{B}_P(\varepsilon).
\label{eq:nominalcontroller}
\ee
However, having a large domain of attraction along with the linear controller \eqref{eq:nominalcontroller} produces very poor local performance, namely, close to the steady state.
To tackle such drawback, we scaled up the nominal control \eqref{eq:nominalcontroller} with $\gamma>1$ and introduced the saturation function proposed in \eqref{eq:def_sat_full} to maintain continuity, invariance and constraint satisfaction.
\end{remark}

\begin{remark}
    It is also noteworthy that the saturation function sat$_\lambda(\bv)$ as in \eqref{eq:def_sat} is Lipschitz continuous due to the convexity of $\mathcal V_c$ given in \eqref{eq:constraint_convex}. This property allows one to practically implement the controller without any chattering, singularities or non-smooth behavior.
\end{remark}

Up to this point, we have proposed a controller ensuring both exponential stability and constraint satisfaction. In the next subsection, we illustrate the synthesis procedure and provide some insight concerning the effect of different parameter values. 
\subsection{Invariant set characterization}
As stated above, in this subsection, we carry out the Procedure \ref{proc:synthesis_control} step by step to illustrate its feasibility and to analyze the parameters' effect.

\begin{figure}[htbp]
    \centering
    \includegraphics[scale=0.5]{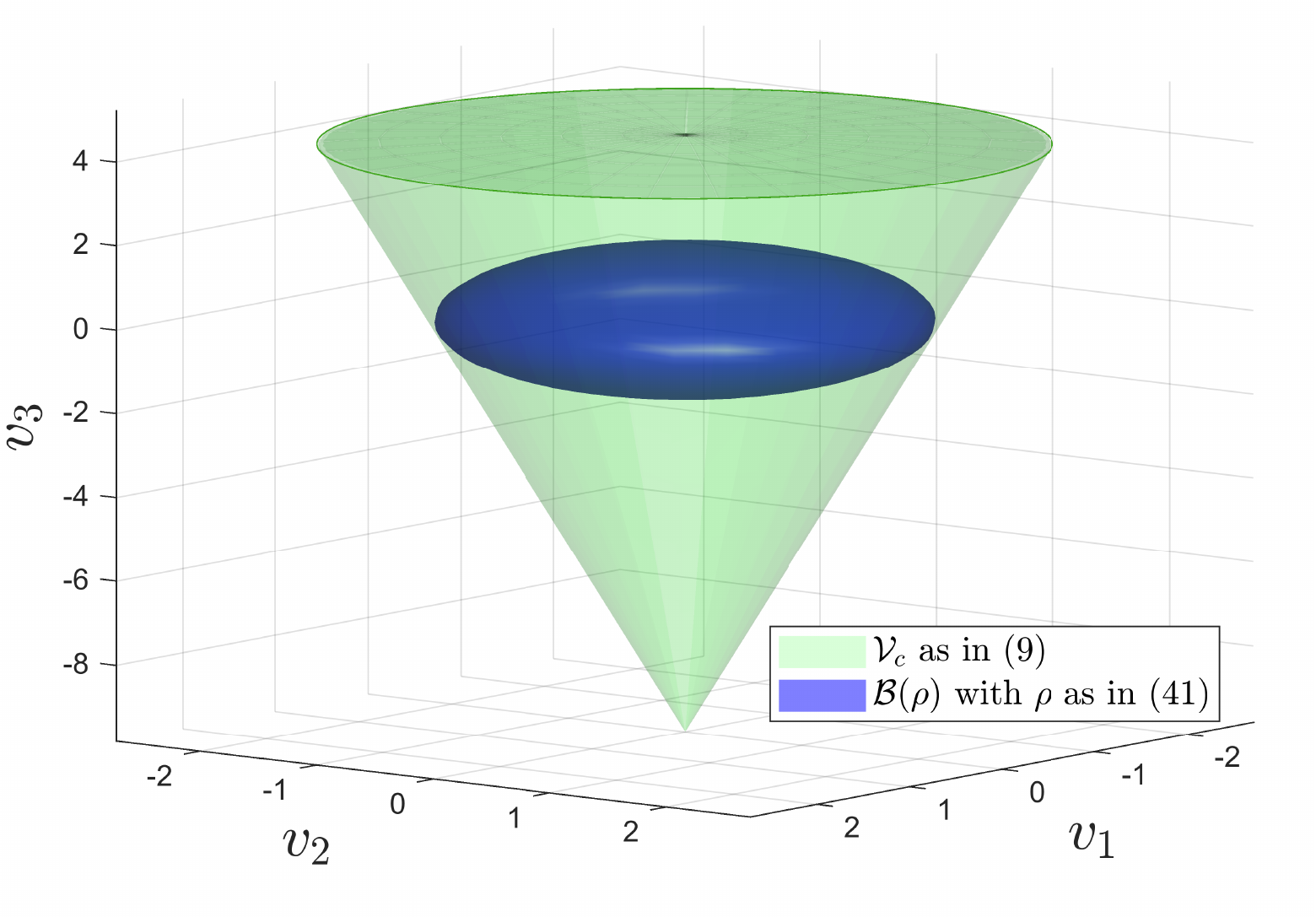}
    \caption{Numerical illustration for the solution of \eqref{eq:ball_in_Vc_rho} - a ball maximally inscribed in $\mathcal{V}_c$.}
    \label{fig:ball_in_Vc_rho}
\end{figure}

With the first step in Procedure \ref{proc:synthesis_control}, instead of empirically choosing a small ball $\brho$ as in \eqref{eq:Ball_in_Vc}, let us propose one candidate solution as follows.
\begin{prop}
The largest ball, of the form \eqref{eq:Ball_in_Vc} and is inscribed in $\mathcal{V}_c$ given in \eqref{eq:constraint_convex}, can be found by solving the following semi-definite programming problem:
\be
\rho^*=\underset{\footnotesize{
\rho }}{\arg \max }\;\rho
\label{eq:ball_in_Vc_rho}
\ee 
\begin{subnumcases}
{\text{s.t }\label{eq:LMI_ball_in_Vc}}
\bbm \bI_{3} & \boldsymbol 0_{3 \times 1} \\ \boldsymbol 0_{1 \times 3} & -\rho \ebm - \tau_1\bbm \bI_{3} &\bar g \\ \bar g^\top & \bar g^\top \bar g -T_{max}^2 \ebm\succeq 0&\\
\bbm \bI_{3} & \boldsymbol 0_{3 \times 1} \\ \boldsymbol 0_{1 \times 3} & -\rho \ebm- \tau_2\bbm H & H^\top \bar g \\ \bar g^\top H & \bar g^\top H \bar g \ebm\succeq 0  &\\
\bbm \bI_{3} & \boldsymbol 0_{3 \times 1} \\ \boldsymbol 0_{1 \times 3} & -\rho \ebm - \tau_3\bbm \boldsymbol 0_{3 \times 3} & -\mu \\ -\mu^\top & -g \ebm\succeq 0&\\
\tau_i\geq 0,i\in\{1,2,3\}
\end{subnumcases}
with $H=\text{diag}(1,1,-\tan\umax\epsilon),\bar g=[0,0,g]^\top$ and $\mu = [0,0,1/2]^\top$.
\QEDA
\end{prop}
\begin{IEEEproof}
Rewriting  the formulation of $\brho$ in \eqref{eq:Ball_in_Vc}, we need the following condition to hold:
\be 
\text{if } \bv\in\brho \Rightarrow \bv \in \mathcal{V}_c
\ee 
or explicitly:
\be
\text{if } \|\bv\|_2^2 \leq \rho \Rightarrow \begin{cases}
(\bv + \bar g)^\top(\bv + \bar g) \leq T_{max}^2  & \\
(\bv + \bar g)^\top H (\bv + \bar g) \leq0 & \\
2\mu^\top \bv  + g\geq 0.&
\end{cases}
\label{eq:ball_in_Vc_cond}
\ee 
Then, by applying the $\mathcal{S}$-procedure as in Proposition \ref{prop:Sproc}, we rewrite \eqref{eq:ball_in_Vc_cond} as in \eqref{eq:LMI_ball_in_Vc}, hence, completing the proof. 
\end{IEEEproof}
An illustration for the optimization problem \eqref{eq:ball_in_Vc_rho} under the condition \eqref{eq:LMI_ball_in_Vc} is provided in Fig. \ref{fig:ball_in_Vc_rho} and Table \ref{tab:First_and_last_Zono_pp}.

Next, to ensure closed-loop stability, it is required to compute a stabilizing matrix $P$ satisfying condition  
\eqref{eq:stabilizing_condition}. By setting $Q=P^{-1}$, pre and post multiplying the result with $Q$, the condition \eqref{eq:stabilizing_condition} yields:
\be 
QA^\top + AQ -2BB^\top\preceq -\alpha Q;\;\;Q\succ 0,
\label{eq:LMI_stabilizingQ}
\ee 
which is, again, an easy-to-handle LMI.
Then finally, with the solution of \eqref{eq:LMI_stabilizingQ}, the invariant set as in \eqref{eq:ellipsoid_PI} is characterized by $P=Q^{-1}$ and $\varepsilon$ found in \eqref{eq:find_ellipsoid}, finalizing the synthesis procedure. 

\begin{figure}[htbp]
    \centering
    \includegraphics[scale=0.46]{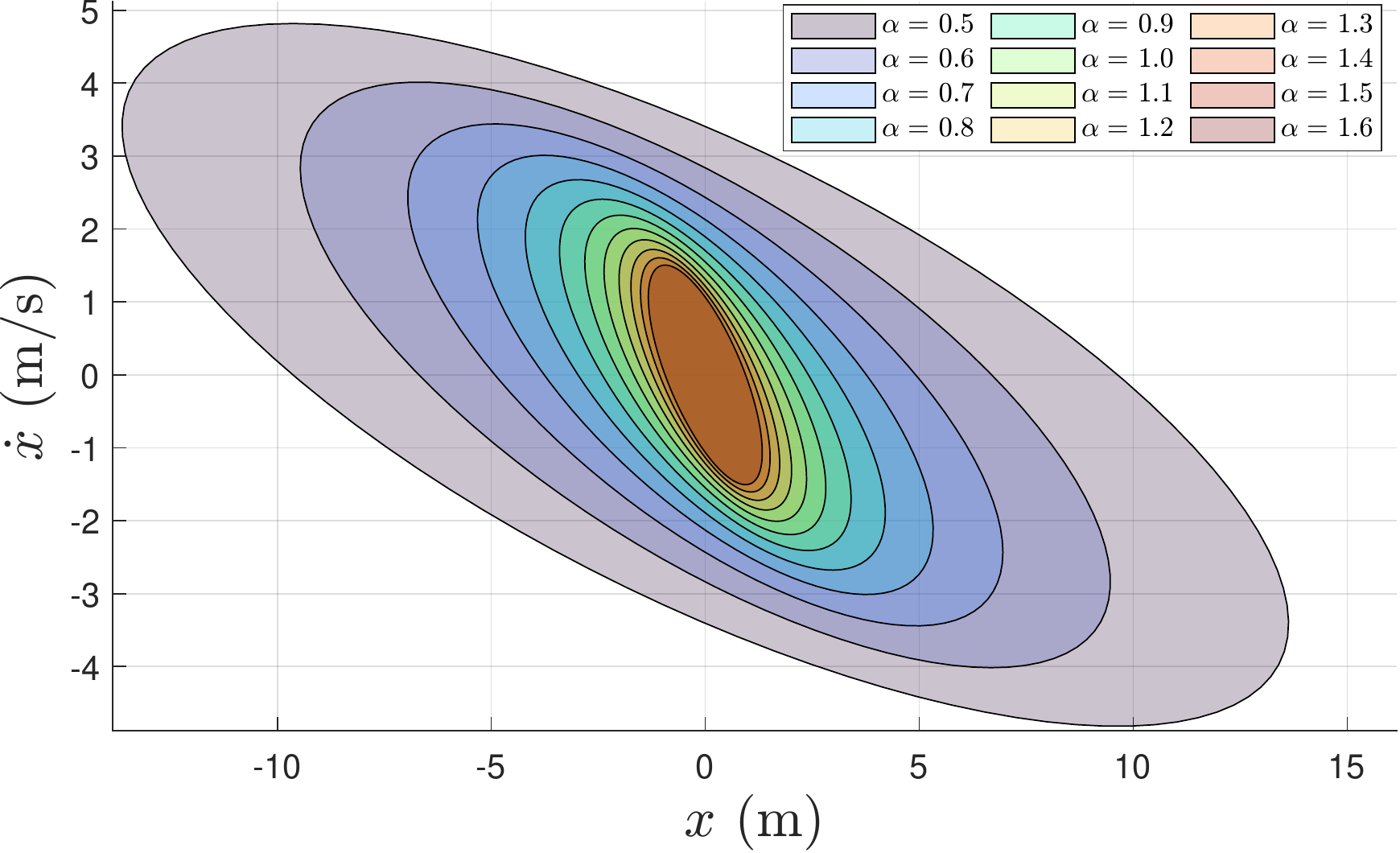}
    \caption{The projection of $\BPep$ into the subspace $(x,\dot x)$ with different choices of $\alpha$.}
    \label{fig:multi_alpha}
\end{figure}

For the sake of illustration, let us depict the simulation result for Procedure \ref{proc:synthesis_control} with different values for $\alpha>0$ (see Fig. \ref{fig:multi_alpha}). 
The numerical results were computed with Yalmip Toolbox \cite{yalmip} together with MATLAB 2021b.
The specifications for the simulation are provided in Table \ref{tab:First_and_last_Zono_pp}.
\begin{table}[htbp]
     \centering
     \caption{Parameters for the invariant set construction}
     \begin{tabular}{|c|c|}
     \hline
         Symbols & Values   \\ \hline \hline
         $g$  &  9.81    $m/s^2$\\ \hline
                $T_{max}$  &   $1.45  g\;m/s^2$   \\ \hline

        $\theta_{max}, \phi_{max}, \umax\epsilon$ as in \eqref{eq:constraint_convex}  &  0.1745  $rad$ $(10^o)$   \\ \hline
        $\rho$ as in \eqref{eq:Ball_in_Vc} &  2.9019\\ \hline
     \end{tabular}
     \label{tab:First_and_last_Zono_pp}
 \end{table}

 As expected, the volume of $\BPep$ increases while $\alpha$ decreases. The downside for this expansion of the invariant ellipsoid is the convergence speed is also reduced, which can be seen from the inequalities \eqref{eq:speed1} and \eqref{eq:speed2}. 

 Meanwhile, the computational effort for the control implementation is the evaluation
of $\lambda^*(\bv)$ as in \eqref{eq:def_sat_opt} when the input is saturated.
One solution is to employ a reliable optimization solver. 
However, taking advantage of the representation of $\mathcal{V}_c$ we can derive an explicit solution as shown in the next section.

\subsection{Implementation algorithm for the saturation function and simulation study}
In view of the explicit computation
of the saturation function, we recall $\text{sat}_\lambda (\bv)$ from \eqref{eq:def_sat_full}:
\begin{subequations}
\begin{align}
 \text{sat}_\lambda (\bv) &=\begin{cases}
 \bv \text{ if } \bv\in \mathcal{V}_c&\\
 \lambda^*(\bv)\bv \text{ if } \bv\notin \mathcal{V}_c.&
 \end{cases}
\label{eq:def_sat_note}
\\
\text{where }\lambda^*(\bv)&=\underset{{
\lambda \bv \in \mathcal{V}_c}}{\arg \max }\;(\lambda ).
\label{eq:compute_lambda}
\end{align}
\end{subequations}
Since \eqref{eq:compute_lambda} is a convex optimization problem, 
the result can be derived directly from the candidate minimizers contained in the following collection $ \mathcal{M}$ (derived from the Karush–Kuhn–Tucker necessary conditions \cite{boyd2004convex}):
\begin{align}
 \mathcal{M}(\bv)&=\Bigg\{ \dfrac{-g}{v_3}, \dfrac{T_{max}}{v_3\sqrt{1+\tan^2(\epsilon_{max})}},\nonumber\\ &\dfrac{-b_1\pm \sqrt{b_1^2-4a_1c_1}}{2a_1}, \dfrac{-b_2\pm \sqrt{b_2^2-4a_2c_2}}{2a_2} \Bigg\}
\label{eq:KKT_M}
\end{align}
with
$$
\begin{aligned}
a_1 &= v_1^2+v_2^2-v_3^2\tan^2(\epsilon_{max}),\,\\b_1 &=-2\tan^2(\epsilon_{max})v_3g,\,c_1=-\tan^2(\epsilon_{max})g^2, \\
a_2&=v_1^2+v_2^2+v_3^2,\,b_2=2v_3g,\,c_2=g^2-T_{max}^2.
\end{aligned}
$$

With this \textit{finite} set, the computation of $\lambda^*(\bv)$ is reduced to:
\be 
\lambda^*(\bv)=\max\mathcal{M}(\bv)\cap (0;1)
\label{eq:def_sat_opt_M}
\ee 
Note that $\mathcal{M}(\bv)$ as in \eqref{eq:KKT_M} may contain complex numbers  that have to be disregarded, since both inputs $\bv$ and $\bu$ are necessarily real. 
Besides,
\eqref{eq:def_sat_opt_M} can be implemented by executing a simple search for a maximum $\lambda$ inside $\mathcal{M}(\bv)\cap \R$ (e.g, a bubble sort).
Briefly, the function \eqref{eq:def_sat_note} (or \eqref{eq:def_sat}) can be represented as the following algorithm:

\begin{algorithm}[H]
\caption{Compute the function $\text{sat}_\lambda (\bv)$ as in \eqref{eq:def_sat_full}.}\label{alg:saturated_func}
\begin{algorithmic}
\REQUIRE $\bv, \umax T, \umax\epsilon$ and $\mathcal{V}_c$ as in \eqref{eq:constraint_convex}.
\ENSURE The result of $\text{sat}_\lambda (\bv)$ as in \eqref{eq:def_sat_full}
\IF{$\bv\notin \mathcal{V}_c$}
\STATE Enumerate 6 components of $\mathcal{M}(\bv)$ as in \eqref{eq:KKT_M};
\STATE $\lambda ^{*} \gets \max\mathcal{M}(\bv)\cap (0;1)$;
\STATE $\bv \gets \lambda^*\bv$;
\ENDIF
\RETURN $\bv$
\end{algorithmic}
\end{algorithm}

With Algorithm \ref{alg:saturated_func}, the control scheme shown in Fig. \ref{fig:scheme_drone} has been fully presented.

\begin{figure}[htbp]
    \centering
    \includegraphics[scale=0.55]{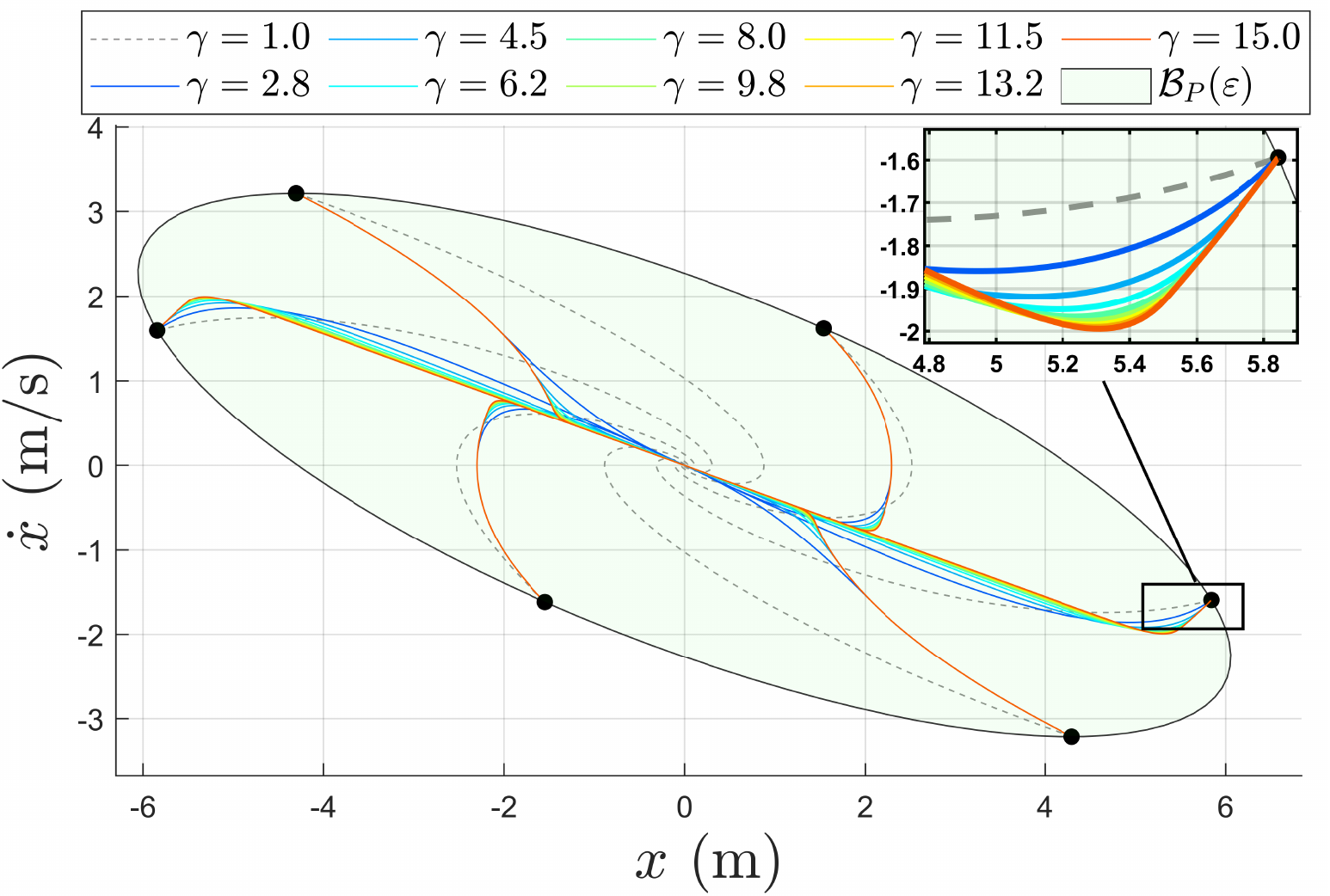}
    \caption{The quadcopter's trajectory with different initial states (black points) implemented with different values of $\gamma$ (projected into $(x,\dot x)$ subspace).}
    \label{fig:TSMC_invariant}
\end{figure}

\emph{Simulation study:} We first illustrate the invariance of $\BPep$ as in \eqref{eq:ellipsoid_PI} regardless of the choice $\gamma\geq 1$, simulating
an origin-tracking problem with different choice for $\gamma$. The initial points are taken on the boundary of $\BPep$. The trajectories' projection\footnote{Similar behaviors can also be found in the $(y,\dot y)$ and $(z,\dot z)$ subspace.} are given in Fig. \ref{fig:TSMC_invariant}. In this simulation, the parameters in Table \ref{tab:First_and_last_Zono_pp} were employed with $\alpha = 0.75$ to solve \eqref{eq:LMI_stabilizingQ}, resulting in an invariant ellipsoid $\BPep$ characterized by $P$ and $\varepsilon$ given in Table \ref{tab:detail_sim1}. 
Moreover, we compare also our Algorithm \ref{alg:saturated_func} with the numerical solution for the saturation function \eqref{eq:def_sat} provided by IPOPT solver\cite{wachter2006implementation} implemented with Yalmip MATLAB Toolbox. In this scenario, the behavior of the system is simulated from the same initial conditions with discretization time as $0.02$ seconds.
\begin{table}[htbp]
    \centering
    \caption{Simulation result for $\BPep$ as in \eqref{eq:ellipsoid_PI}}
    \begin{tabular}{|c|c|}
    \hline
        $ \varepsilon$ as in \eqref{eq:ellipsoid_PI}&  3.8692 \\ \hline
        $P=Q^{-1}$ as in \eqref{eq:LMI_stabilizingQ}&  $\bbm 0.2109\boldsymbol I_{3} & 0.2813\boldsymbol I_{ 3 }\\ 0.2813\boldsymbol I_{ 3 } & 0.7500\boldsymbol I_{ 3 } \ebm $ \\\hline
    \end{tabular}
    \label{tab:detail_sim1}
\end{table}

\begin{figure}[htbp]
    \centering
    \includegraphics[scale=0.5]{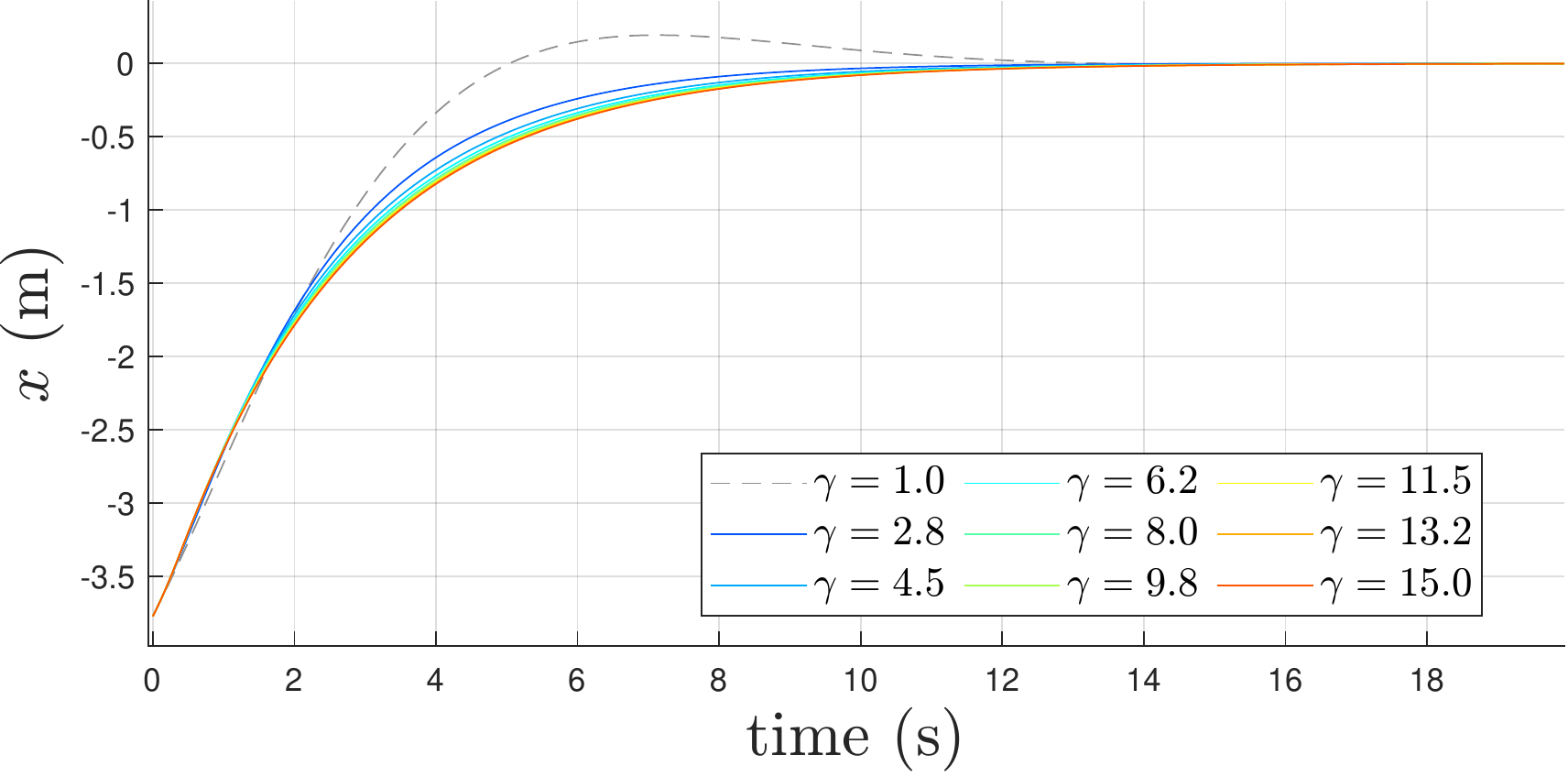}
    \caption{Simulated trajectories in time with different values of $\gamma$.}
    \label{fig:TSMC_x}
\end{figure}

\begin{figure}[htbp]
    \centering
    \includegraphics[scale=0.5]{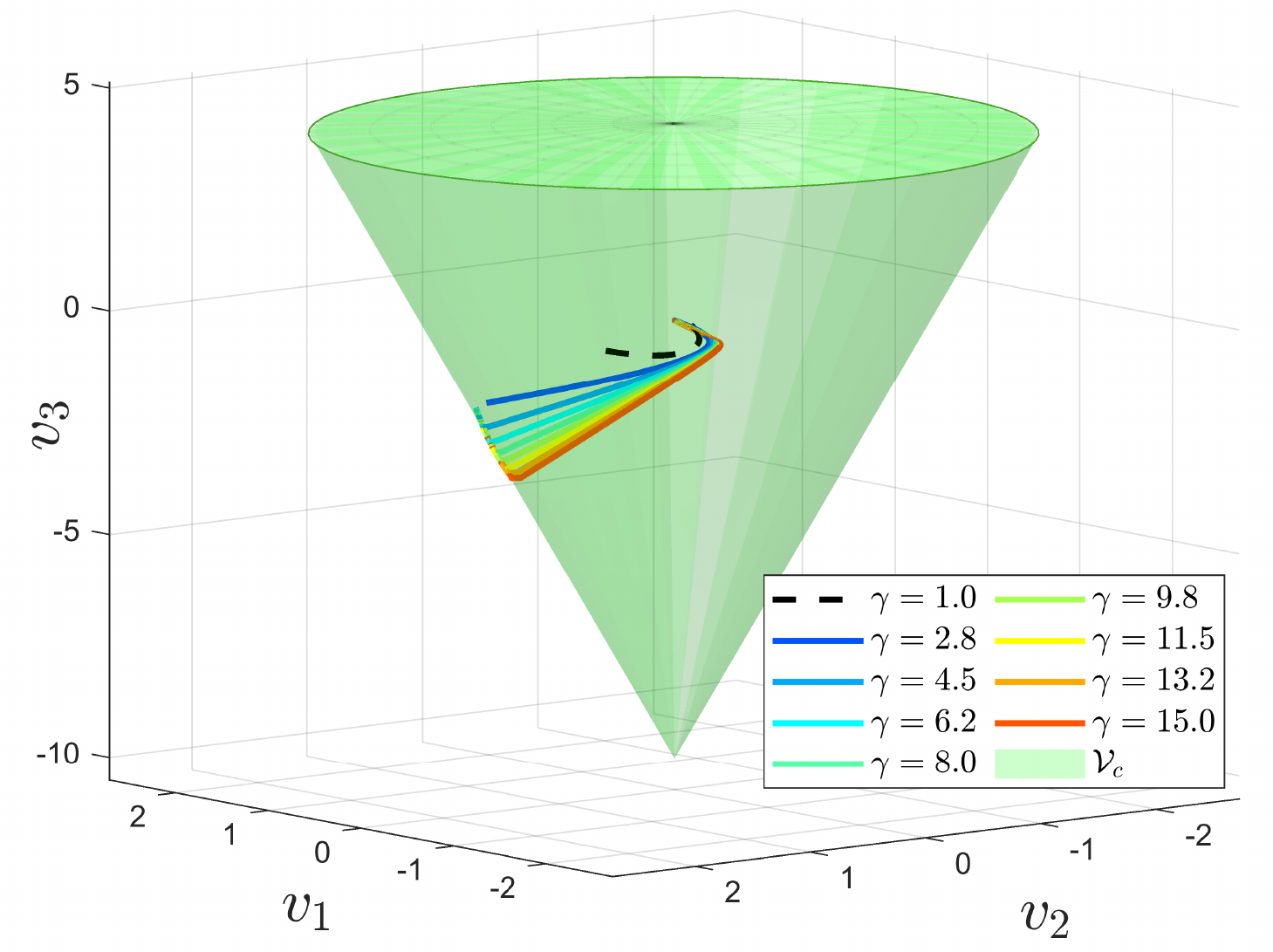}
   \caption{Saturated input $\bv$ simulated with different values of $\gamma$.}
    \label{fig:TSMC_cone_sat}
\end{figure}

As can be seen from Fig. \ref{fig:TSMC_invariant}, the control guarantees the invariance of the set $\BPep$. The same confirmation can be made in all other subspaces during the simulation.
Moreover, the accuracy of Algorithm \ref{alg:saturated_func} can also be seen in Fig. \ref{fig:TSMC_cone_sat} where the new input $\bv$ is ``saturated" on the surface of $\mathcal{V}_c$.
Finally, the effect of the choice for $\gamma$ can also be distinguished from both Fig. \ref{fig:TSMC_x} and Fig. \ref{fig:TSMC_cone_sat} with the initial state $\bxi(0) = [ -3.77,
   -0.46,
   -3.60,
    0.94,
   -0.24,
    2.31]^\top$. More specifically, when $\gamma=1$, the controller is never saturated due to the property \eqref{eq:condition_lambdastart} as proven before. At the same time, when $\gamma>1$, the damping effect appears to be increased together with $\gamma$. However, as can be seen from Fig. \ref{fig:TSMC_x}, choosing a larger value for $\gamma$ does not necessarily imply a faster convergence, hence, giving us an indication not to abusively augment the value of this scalar while designing the controller.

\begin{figure}[htbp]
    \centering
    \includegraphics[scale=0.5]{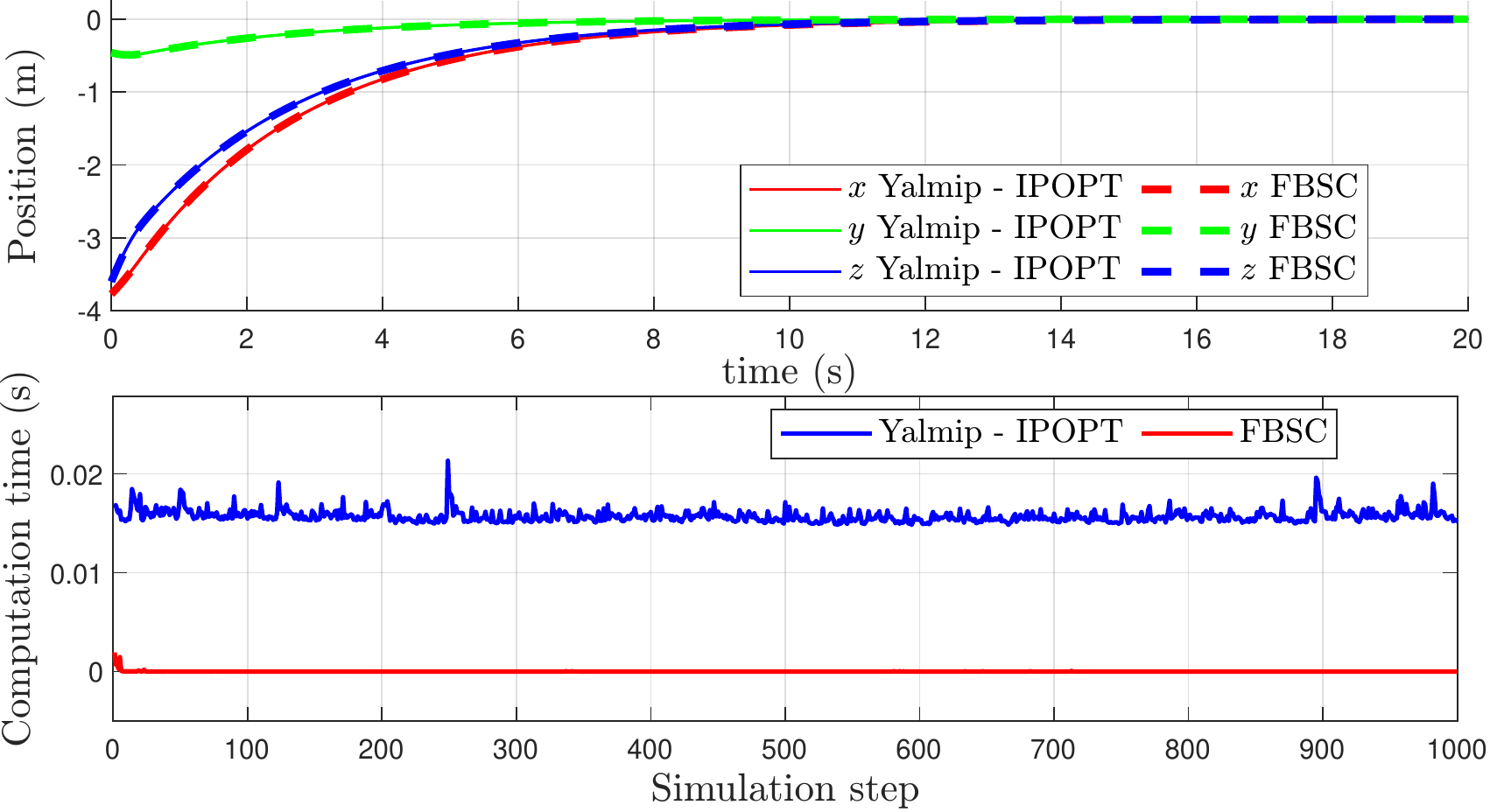}
    \caption{The quadcopter's trajectories and computation time comparison with the IPOPT solver ($\gamma=15$).}
    \label{fig:compare_2med}
\end{figure}

Furthermore, it is noticeable from Fig. \ref{fig:compare_2med} that although Algorithm \ref{alg:saturated_func} generates the same trajectory as the IPOPT solver does, the required computation time for our solution is significantly lower compared to that of the solver.
This observation hence, again, confirms not only the validity but also the computational advantage of our work.

In the next section, the controller will be validated via experiments.
\section{Experimental validation}
\label{sec:experiments}
To illustrate the applicability of our controller and highlight its advantages, we carry out experimental tests
and compare them
with the results of existing methods in the literature. 

\subsection{Experimental setup and scenarios}
For the experiments, we use a
nano-drone Crazyflie 2.1 platform. This aerial drone provides us the access to its high level controller, which requires exactly the input $\bu=[T,\phi,\theta]^\top$ as in \eqref{eq:drone_dyna} together with the desired angular rate for the yaw angle ($\psi$) (for our scenario, this variable $\reff{\dot\psi}$ is considered 0). The packet containing $\bu$ and $\reff{\dot\psi}=0$ will be computed via the proposed controller on a ground station computer and sent to the Crazyflie via a PA USB radio dongle. 
Note that, before being sent to the drone, the normalized thrust $T\, (m/s^2)$ needs to be converted to the thrust unit of the Crazyflie platform, ranging from 
 0 to 65535. Details on the conversion  can be found on the documentation of Bitcraze AB team with the system identification fitting model\cite{CF_thrust}. 
 Furthermore, the implementation is carried out using the \textit{Commander} class provided in the API reference for CFLib\cite{CF_commander} with Python 3.9.
 
For the feedback sensors, we employ a state-of-the-art indoor optical motion capture system. More specifically, 8 infrared cameras  from Qualisys \cite{senior2004qualisys} are used to capture the position and orientation of a \textit{rigid body} identified with a collection of reflective markers which are glued on the Crazyflie quadcopter. Then, based on the positions of the markers captured by the cameras, the motion of the drone are estimated by \textit{Qualisys Track Manager} software and sent to the ground station computer in real time via TCP/IP. The cameras' capture rate is chosen at 120Hz for this experiment,
 while the controller's sampling time is chosen to be 0.075(sec).  

To examine the performance of the controller \eqref{eq:control_sat}, we introduce the following tracking scenarios.
\begin{itemize}
    \item {Ref. 1:} In this scenario, with a fixed choice of $P=Q^{-1}$ satisfying \eqref{eq:LMI_stabilizingQ}, the proposed controller will be employed to track a stationary point $\reff\bxi=[0.3,0.3,0.8,0,0,0]^\top$. We consider various values for $\gamma$ to analyze the impact of such constant on the quadcopter's tracking performance in real implementation.
    \item Ref. 2: For comparison, we adopt the time varying reference used for validation in \cite{santoso2019hybrid} where the authors used  a hybrid PID-fuzzy controller to stabilize the linearly approximated system around its equilibrium point.
    \item Ref. 3: We also consider the smooth trajectory generated via B-spline parameterization \cite{prodan2019necessary,do2021analysis,thinhECC23}, and tracked with a model predictive controller.
    \item Ref. 4: Furthermore, for comparison, we select also the circular trajectory applied in \cite{greeff2018flatness} where the authors applied the model predictive control for the linear system \eqref{eq:linearized_drone} in the flat output space but with a conservative box-type
    constraints as opposed to $\mathcal{V}_c$ as in \eqref{eq:constraint_convex}. For this last scenario, the flat output given in \eqref{eq:flat_output_def} is parameterized as:
    $$
\begin{aligned}
&\sigma_{1}^\mathrm{ref}(t)=0.5\cos\omega t+0.2; \,\sigma_{2}^\mathrm{ref}(t)=0.5\sin\omega t\\
&\sigma_{3}^\mathrm{ref}(t)=0.3(m);\,\omega=0.3\pi.
\end{aligned}
$$
\end{itemize}

Finally, in all the scenarios, we simply employ the stabilizing matrix $P=Q^{-1}=\bbm 0.9766\boldsymbol I_{ 3 } & 0.7813\boldsymbol I_{3 }\\ 0.7813\boldsymbol I_{3} & 1.2500\boldsymbol I_{ 3 } \ebm $ for the controller \eqref{eq:control_sat} as a solution of \eqref{eq:LMI_stabilizingQ} while choosing $\alpha=1.25$, so that the resulted invariant set $\BPep$ as in \eqref{eq:ellipsoid_PI} covers the drone's initial position.
The video
 of the experiment is available at \url{\videourl}.
 
\subsection{Experimental results and discussion}
\begin{figure}[htbp]
    \centering
    \includegraphics[scale=0.5]{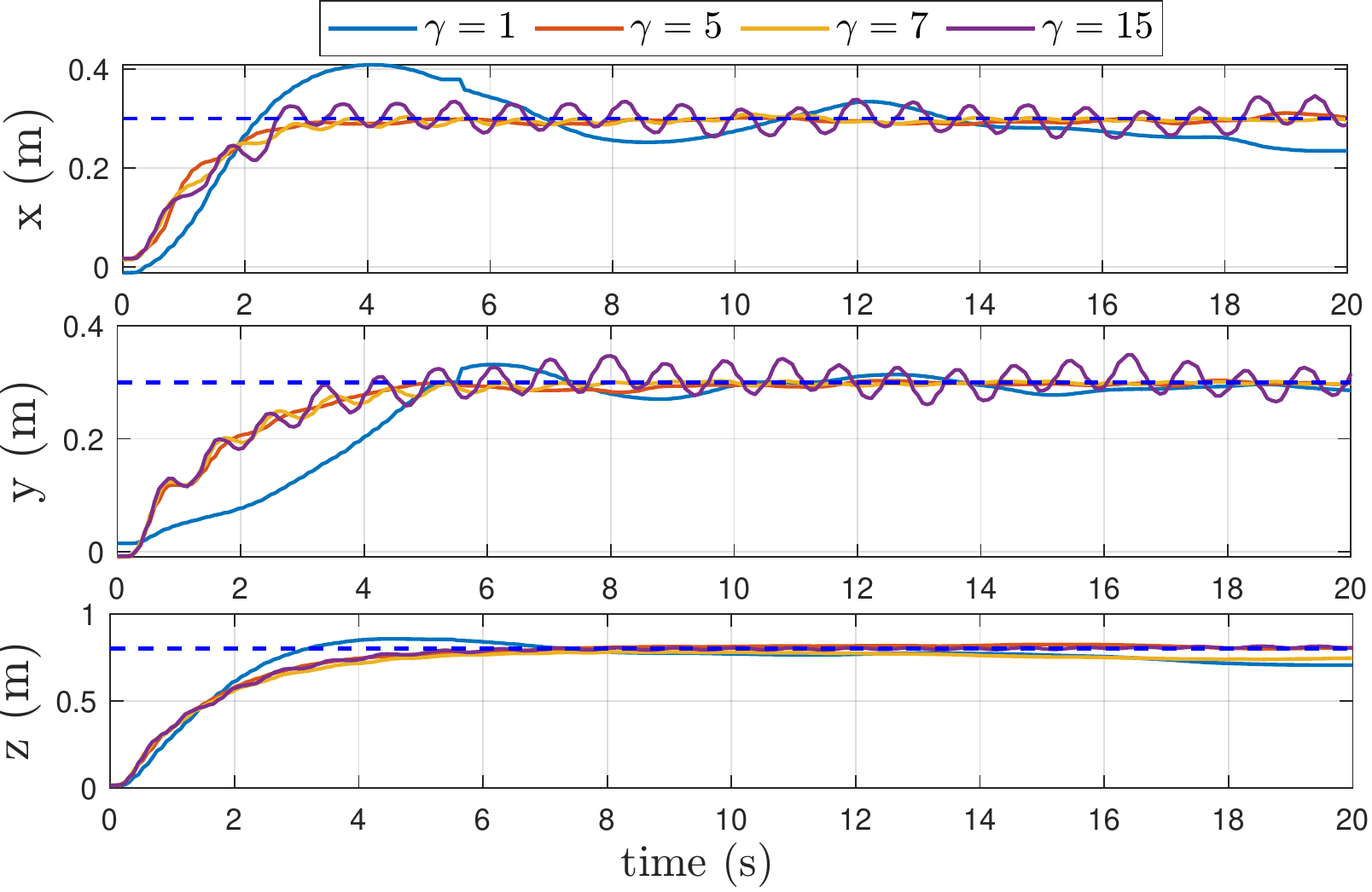}
    \caption{Ref. 1 (blue-dashed line) tracked with different values of $\gamma$.}
    \label{fig:setpoint_multigamma}
\end{figure}

Overall, as can be seen in Fig. \ref{fig:setpoint_multigamma}, Fig \ref{fig:trajtrack_Ref234} and Table \ref{tab:spec_exp}, all four trajectories were tracked properly with the root-mean-square (RMS) of the tracking errors under 10 centimeters. Moreover, it can be observed that the input constraints are always respected in all scenarios and the computation time is remarkably small compared to the other constraint-handling controllers with stability guarantees found in the literature \cite{greeff2018flatness,nguyen2020stability,nguyen2021stabilizing}. 
Further details on the experimental results and our proposed controller performance are given in the following.

\begin{figure}[htbp]
    \centering
    \includegraphics[scale=0.45]{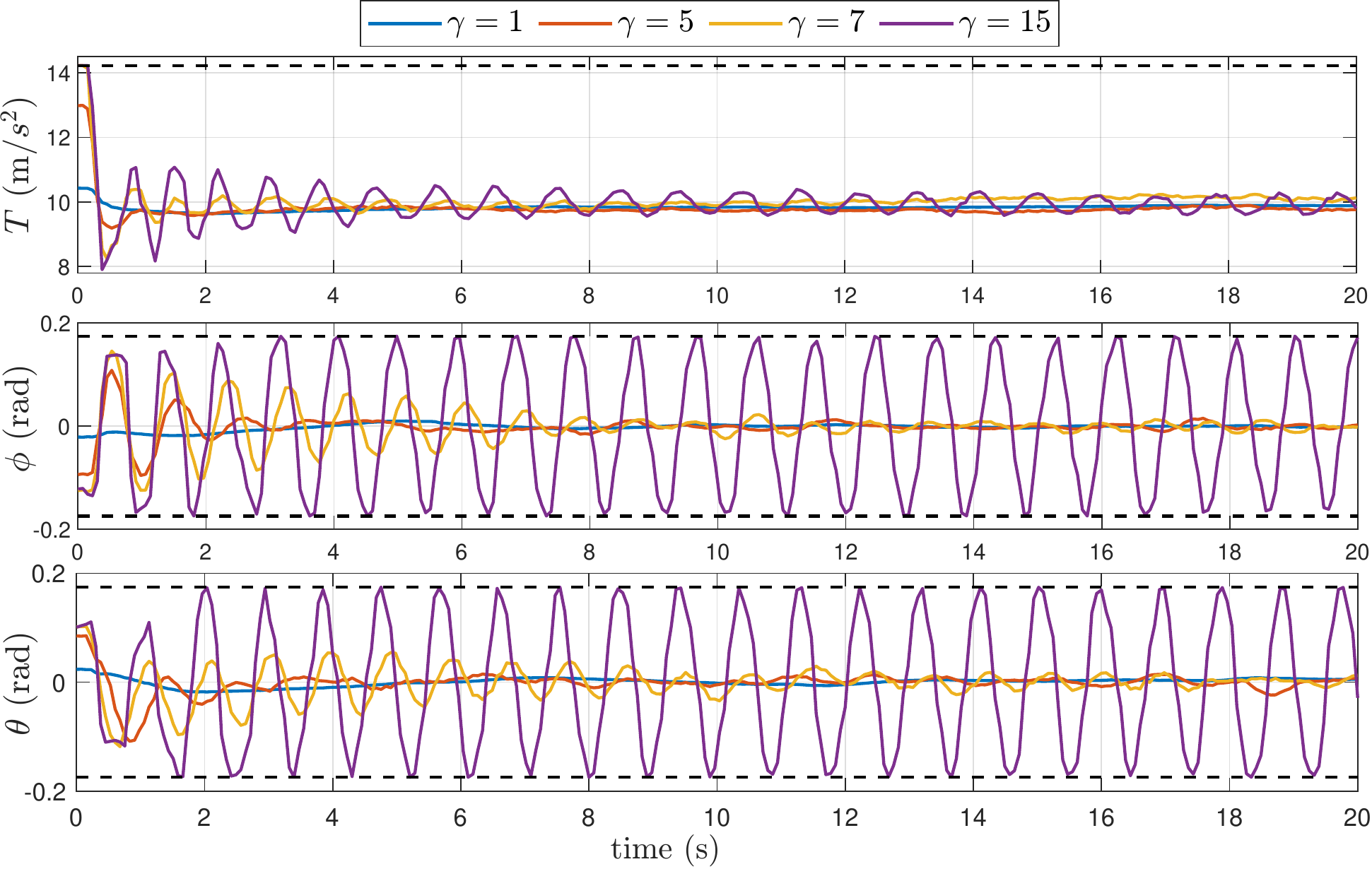}
    \caption{Input $\bu$ with different values of $\gamma$ with the set-point tracking problem and its bounds $\mathcal{U}$ as in \eqref{eq:orginal_input_constr} (black dashed line).}
    \label{fig:input_setpoint_multigamma}
\end{figure}

\begin{figure}[htbp]
    \centering
    \includegraphics[scale=0.4]{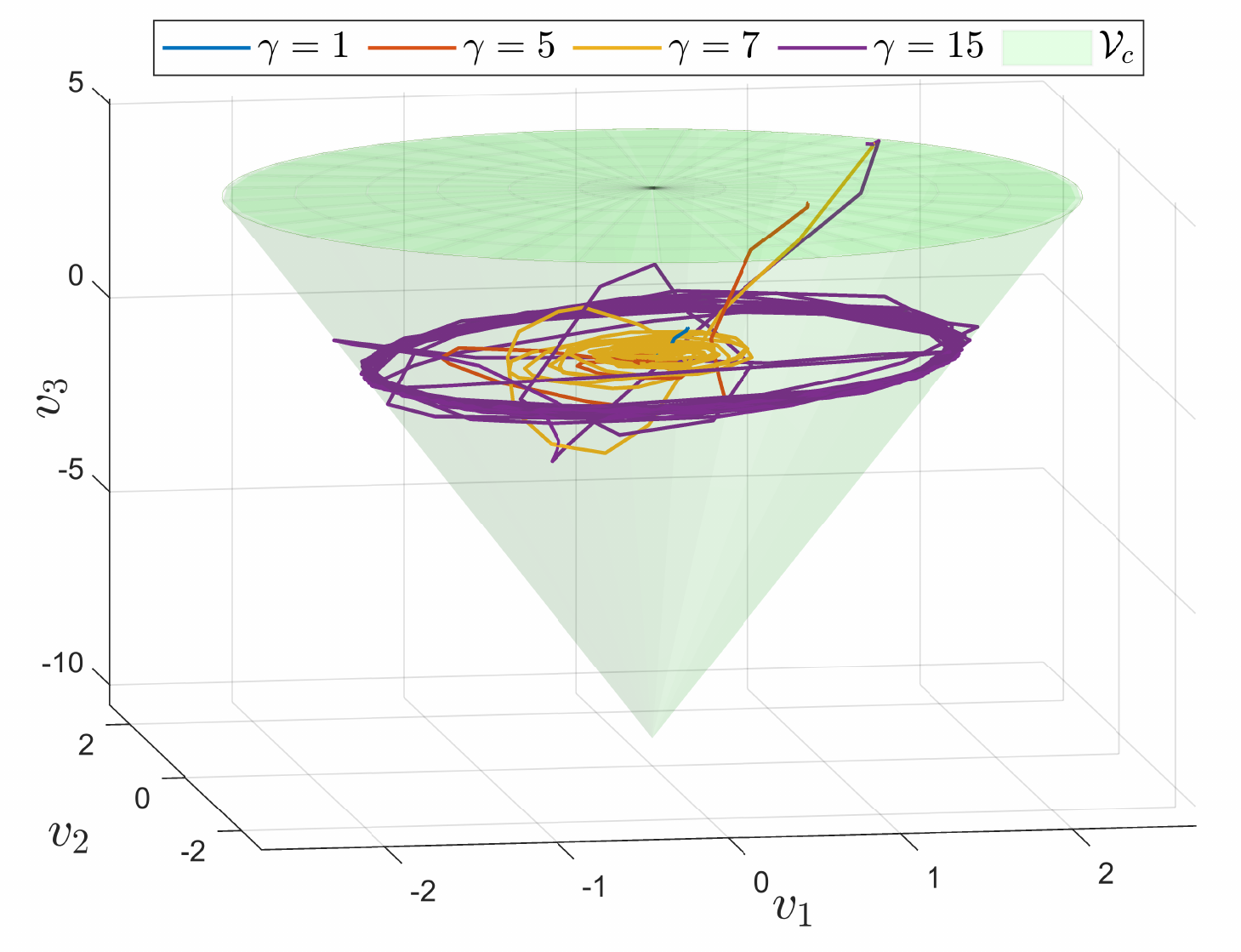}
    \caption{Input $\bv$ in the flat output space, with different values of $\gamma$ in Ref. 1 tracking problem and its constraint set $\mathcal{V}_c$ as in \eqref{eq:constraint_convex} .}
    \label{fig:virtual_input_setpoint_multigamma}
\end{figure}

Firstly, the tracking result for Ref. 1 is depicted in Fig. \ref{fig:setpoint_multigamma} while the corresponding input signals in the original space and in the flat output space are given in Fig. \ref{fig:input_setpoint_multigamma} and \ref{fig:virtual_input_setpoint_multigamma}, respectively. As expected from the simulation in Fig. \ref{fig:TSMC_x}, after increasing to a certain value of $\gamma>1$ in \eqref{eq:control_sat}, not only does the settling time not decrease, but also an oscillating effect appears (Fig. \ref{fig:setpoint_multigamma}). While this effect is not apparent in the simulation, it can be explained as the outcome of the high gain controller that awakes the parasitic dynamics existing
within the vehicles (e.g. the rotational/attitude dynamics \cite{hua2009control}). Therefore, it is practically verified that $\gamma$ does not need to be chosen excessively large to achieve a favorable performance. Besides, from Fig. \ref{fig:virtual_input_setpoint_multigamma}, the effectiveness of the proposed saturation function (computed with Algorithm \ref{alg:saturated_func}) is again validated via the constrained input $\bv\in \mathcal{V}_c$ in \eqref{eq:constraint_convex}.

Next, it is also notable that the average computation time rests approximately at 0.1 milliseconds (see Table \ref{tab:spec_exp}). In terms of computing time, this result surpasses previously mentioned  controllers in the literature (around 5 to 50 milliseconds and above), which also guarantee stability and constraint satisfaction \cite{greeff2018flatness,nguyen2020stability,nguyen2021stabilizing}. This satisfactory performance comes from the fact that we provide an explicit formula of the controller from computing the feedback input to saturating such signal via Algorithm \ref{alg:saturated_func}. 
Although in this experiment, all the control computation is executed remotely in the ground station computer, 
these results appear promising
for embedding the controller directly on the drone's microcontroller thanks to the simplicity of the proposed scheme.
\begin{figure}[htbp]
    \centering
    \includegraphics[scale=0.5]{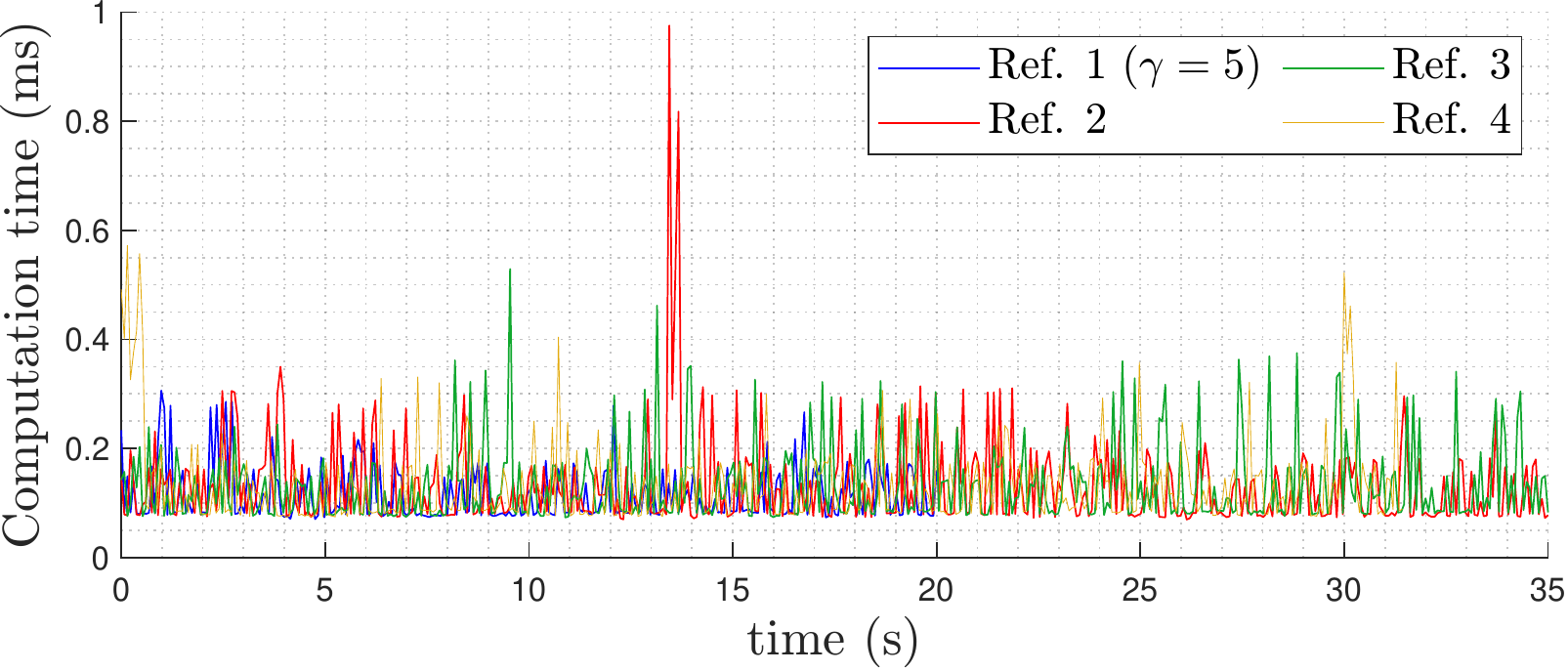}
    \caption{Computation time with the proposed controller.}
    \label{fig:comptime_Ref234}
\end{figure}
\begin{figure}[htbp]
    \centering
\includegraphics[scale=0.525]{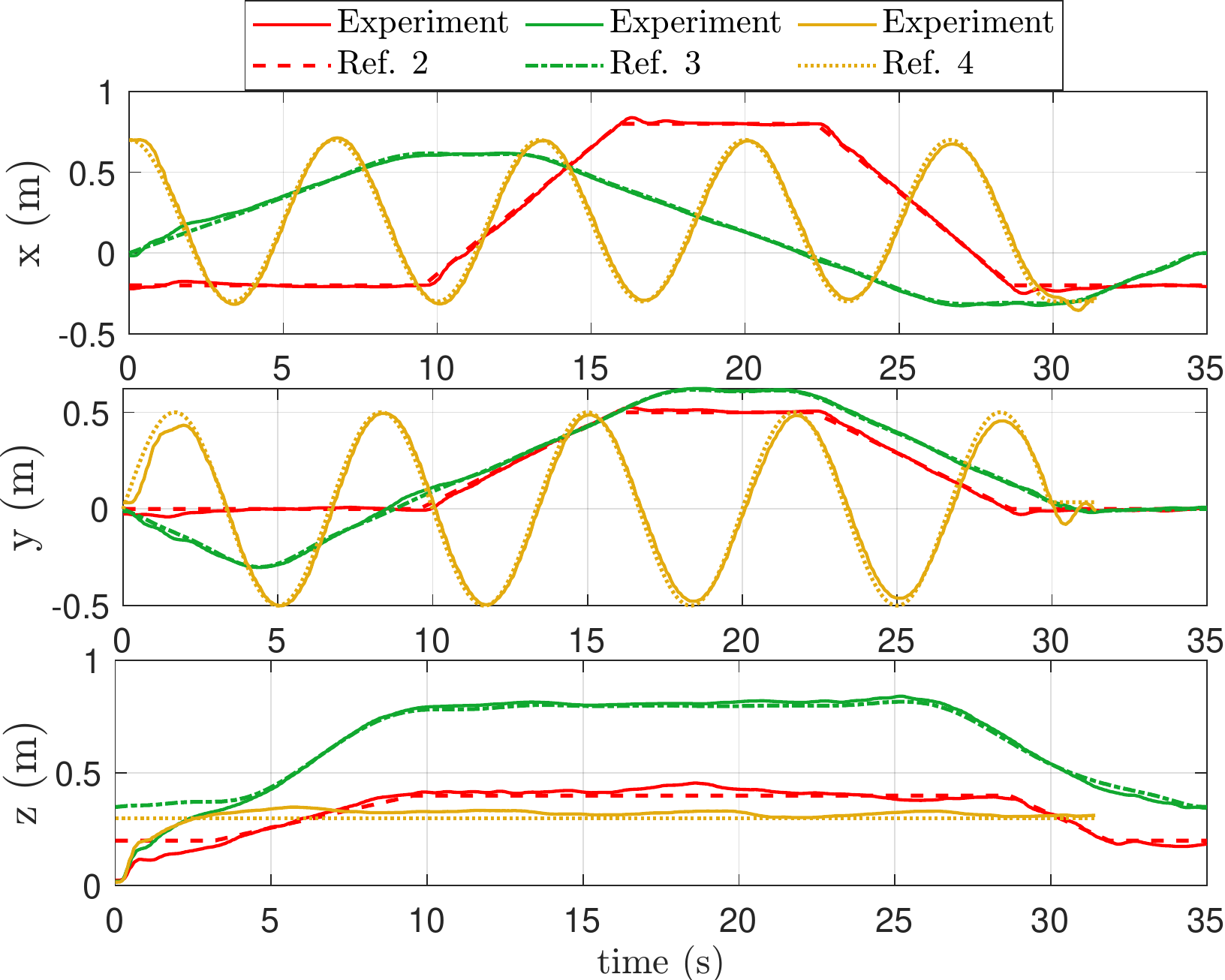}
    \caption{Position tracking with different trajectories.}
    \label{fig:trajtrack_Ref234}
\end{figure}

\begin{figure}[htbp]
    \centering
    \includegraphics[scale=0.5]{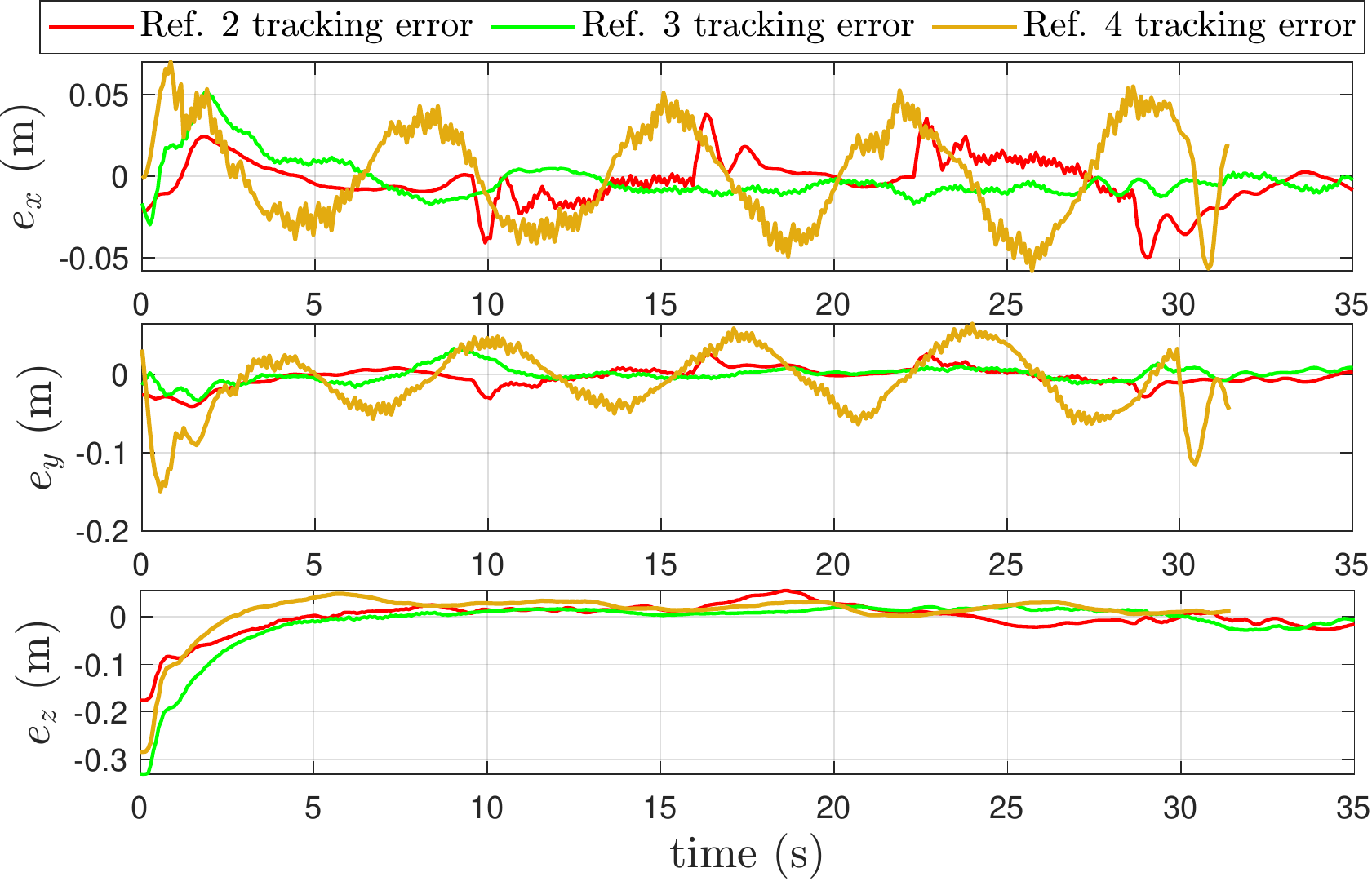}
    \caption{Tracking error $e_q=q-\reff q,q\in\{x,y,z\}$ with different trajectories.}
    \label{fig:error_Ref234}
\end{figure}

\begin{figure}[htbp]
    \centering
    \includegraphics[scale=0.5]{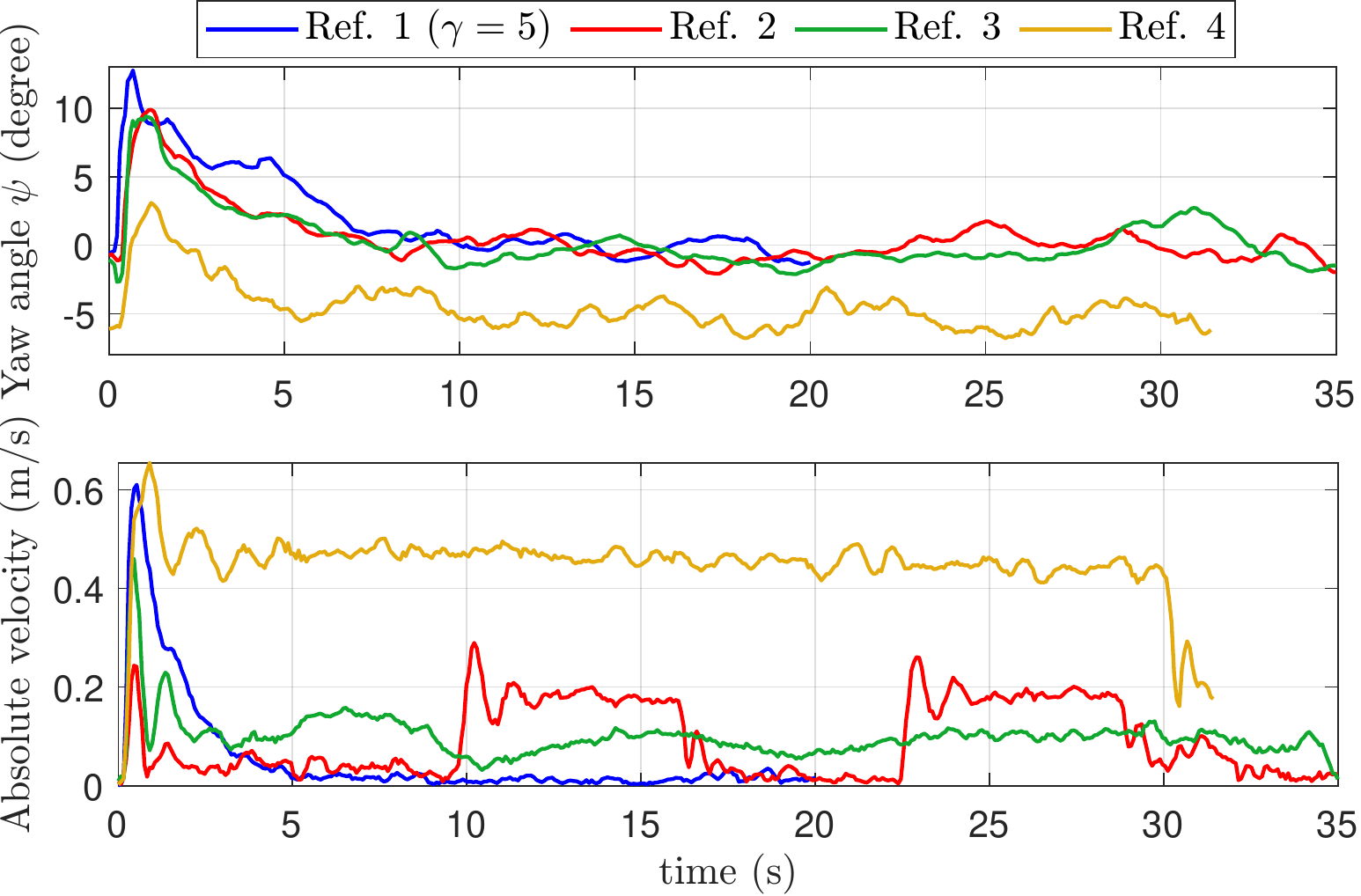}
    \caption{The yaw angle ($\psi$) and the absolute velocity of the quadcopter ($\sqrt{\dot x^2+\dot y^2+\dot z^2}$) during the experiments.}
    \label{fig:Yaws_vel}
\end{figure}

Finally, regarding the closed-loop performance, the tracking errors provided by the proposed method can be considered commensurate to the flatness-based MPC, the well-known piece-wise affine MPC (where the model is approximated along the trajectory), or the approximation-dependent fuzzy-based method proposed in \cite{thinhECC23}, Fig. 29 in \cite{santoso2019hybrid}. 

\begin{figure}[htbp]
    \centering
    \includegraphics[scale=0.5]{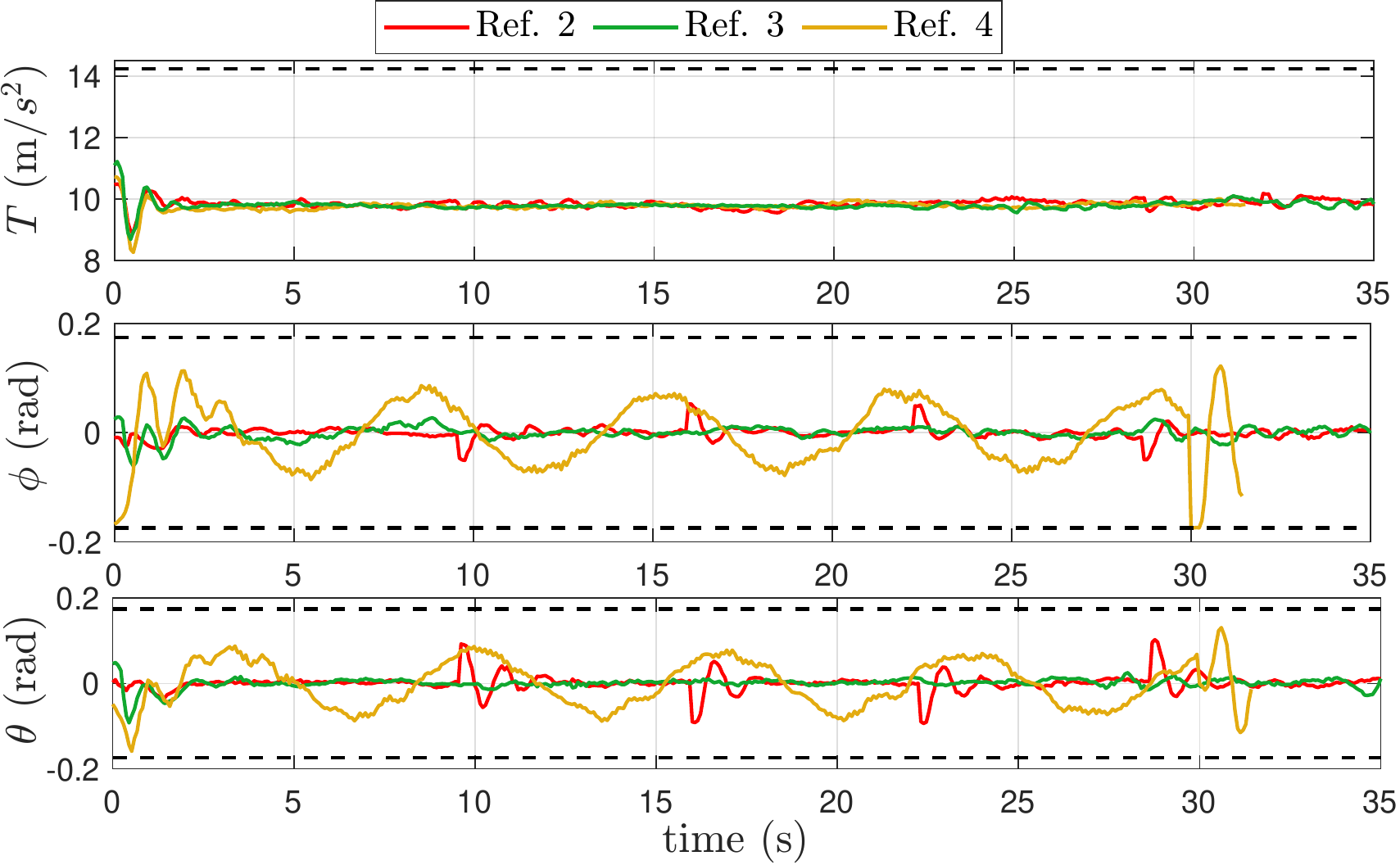}
    \caption{Input signal $\bu$ in the tracking problem for Ref.2, Ref.3 and Ref.4.}
    \label{fig:real_input_Ref234}
\end{figure}
\begin{table}[htbp]
  \centering
  \caption{Experiment specifications and results}
    \begin{tabular}{|l|p{0.7cm}|p{0.7cm}|p{0.7cm}|p{0.7cm}|}
    \hline
         & Ref. 1  & Ref. 2  & Ref. 3 &Ref. 4\\
    \hline
    \multicolumn{1}{|p{11.335em}|}{Average computation time\newline{}(s) $\times 10^{-4}$} & \multirow{2}[1]{*}{$1.164$} &\multirow{2}[1]{*}{$ 1.321$}  &\multirow{2}[1]{*}{$1.327$} &\multirow{2}[1]{*}{$1.323$} \\
    \hline
    RMS  of tracking error (cm) & 10.03 & 1.99 & 2.57& 3.89\\
    \hline
    $\gamma$ as in \eqref{eq:control_sat}& 5 &4.5&4.5& 4.5
    \\
    \hline
    $P$ as in \eqref{eq:control_sat} & \multicolumn{4}{c|}{$\bbm 0.9766\boldsymbol I_{3 } & 0.7813\boldsymbol I_{3 }\\ 0.7813\boldsymbol I_{3 } & 1.2500\boldsymbol I_{ 3 } \ebm $} 
    \\ \hline
Sampling time (s) & \multicolumn{4}{c|}{0.075} 
    \\ \hline
    \end{tabular}%
  \label{tab:spec_exp}%
\end{table}%

\section{Conclusion}
\label{sec:conclude}

In this work, an explicit saturated controller for the quadcopter system was presented with the guarantee for both constraint satisfaction and its stability. Its practical viability was also validated in both simulations and experiments.
The controller was constructed by taking advantage of the flatness-based linearization of the system and its corresponding distorted input constraints in the new coordinates. 
Hence, its validated applicability not only opens the door to further developments on the computationally low-cost controllers for quadcopter vehicles, but also
the analogous application on other classes of differentially flat systems.


\end{document}